\documentclass[sigconf]{acmart}

\copyrightyear{2024}
\acmYear{2024}
\setcopyright{acmlicensed}\acmConference[ICSE '24]{2024 IEEE/ACM 46th International Conference on Software Engineering}{April 14--20, 2024}{Lisbon, Portugal}
\acmBooktitle{2024 IEEE/ACM 46th International Conference on Software Engineering (ICSE '24), April 14--20, 2024, Lisbon, Portugal}
\acmDOI{10.1145/3597503.3639184}
\acmISBN{979-8-4007-0217-4/24/04}

\AtBeginDocument{%
  \providecommand\BibTeX{{%
    Bib\TeX}}}

\AtBeginDocument{%
	\providecommand\BibTeX{{%
			\normalfont B\kern-0.5em{\scshape i\kern-0.25em b}\kern-0.8em\TeX}}}


\newcommand{\name}{PyTy}
\newcommand{\dataset}{PyTyDefects}
\newcommand{\pr}[1]{20}

\newcommand{\code}[1]{\texttt{\small#1}}

\usepackage{booktabs}
\usepackage{amsmath}
\usepackage{listings}
\usepackage{xcolor}
\definecolor{celadon}{RGB}{221, 250, 218}
\makeatletter
\newcommand\HL{%
   \gdef\lst@alloverstyle##1{%
     \fboxrule=0pt
     \fboxsep=0pt
     \colorbox{celadon}{\strut##1}%
   }%
}

\newcommand\HLoff{%
   \xdef\lst@alloverstyle##1{##1}%
}
\definecolor{shadecolor}{RGB}{235,235,235}

\newcommand\HLL{%
	\gdef\lst@alloverstyle##1{%
		\fboxrule=0pt
		\fboxsep=0pt
		\colorbox{shadecolor}{\strut##1}%
	}%
}

\newcommand\HLLoff{%
	\xdef\lst@alloverstyle##1{##1}%
}

\definecolor{reedcolor}{RGB}{255,204,203}

\newcommand\HLLL{%
	\gdef\lst@alloverstyle##1{%
		\fboxrule=0pt
		\fboxsep=0pt
		\colorbox{reedcolor}{\strut##1}%
	}%
}

\newcommand\HLLLoff{%
	\xdef\lst@alloverstyle##1{##1}%
}

\usepackage{hyperref}
\usepackage{cleveref}
\usepackage{algorithm}
\usepackage{algorithmicx}
\usepackage[noend]{algpseudocode}
\usepackage{syntax}
\usepackage{subcaption}
\usepackage{multirow}
\usepackage{graphicx}
\usepackage{xcolor}
\usepackage{fancyvrb}
\usepackage{ragged2e}
\usepackage{array}
\usepackage{enumitem}
\setlist[itemize]{leftmargin=*}
\setlist[enumerate]{leftmargin=*}

\usepackage{tcolorbox}
\newenvironment{finding}[1]{
	\begin{tcolorbox}[colback=blue!5!white,colframe=blue!5!white,arc=0mm,grow to left by=0mm,left=0mm,grow to right by=0mm,left=0mm,right=0mm,top=0mm,bottom=0mm]
		\textbf{Answer to #1:}
	}
	{
	\end{tcolorbox}
}

\newcommand\vartextvisiblespace[1][.4em]{%
	\mbox{\kern.2em\vrule height.3ex}%
	\vbox{\hrule width#1}%
	\hbox{\vrule height.3ex}
	\mbox{\kern.2em}%
}

\begin{document}
	
	\title[\name{}: Repairing Static Type Errors in Python]{\name{}: Repairing Static Type Errors in Python}
	
	\author{Yiu Wai Chow}
	\affiliation{
		\institution{University of Stuttgart}
		\country{Germany}
	}
	\email{victorcwai@gmail.com}
	
	\author{Luca Di Grazia}
	\affiliation{
		\institution{University of Stuttgart}
		\country{Germany}
	}
	\email{work@lucadigrazia.com}
	
	\author{Michael Pradel}
	\affiliation{
		\institution{University of Stuttgart}
		\country{Germany}
	}
	\email{michael@binaervarianz.de}
	\begin{abstract}
		Gradual typing enables developers to annotate types of their own choosing, offering a flexible middle ground between no type annotations and a fully statically typed language.
As more and more code bases get type-annotated, static type checkers detect an increasingly large number of type errors.
Unfortunately, fixing these errors requires manual effort, hampering the adoption of gradual typing in practice.
This paper presents \name{}, an automated program repair approach targeted at statically detectable type errors in Python. 
The problem of repairing type errors deserves specific attention because
it exposes particular repair patterns,
offers a warning message with hints about where and how to apply a fix, and because gradual type checking serves as an automatic way to validate fixes.
We addresses this problem through three contributions:
(i) an empirical study that investigates how developers fix Python type errors, showing a diverse set of fixing strategies with some recurring patterns;
(ii) an approach to automatically extract type error fixes, which enables us to create a dataset of 2,766 error-fix pairs from 176 GitHub repositories, named \dataset;
(iii) the first learning-based repair technique for fixing type errors in Python.
Motivated by the relative data scarcity of the problem, the neural model at the core of \name{} is trained via cross-lingual transfer learning.
Our evaluation shows that \name{} offers fixes for ten frequent categories of type errors, successfully addressing 85.4\% of 281 real-world errors.
This effectiveness outperforms state-of-the-art large language models asked to repair type errors (by 2.1x) and complements a previous technique aimed at type errors that manifest at runtime.
Finally, \pr{} out of 30 pull requests with \name{}-suggested fixes have been merged by developers, showing the usefulness of \name{} in practice.

	\end{abstract}

	
	\maketitle
	
	\section{introduction}
\label{sec:intro}
Dynamically typed languages, such as Python and JavaScript, have become very popular.\footnote{\url{https://octoverse.github.com/\#top-languages-over-the-years}} 
One reason is their lightweight syntax, which does not require developers to specify types for parameters, return values, or variables. Because this flexibility may negatively affect the maintainability and robustness of code, in 2015, Python adopted optional type annotations, enabling developers to annotate types of their choosing.

\paragraph{Context} Since their introduction into the Python language, type annotations have been getting increasingly popular~\cite{Digrazia2022}.
To support developers, several automated approaches for adding type annotations to existing code bases have been proposed, e.g., TypeWriter~\cite{fse2020}, DeepTyper~\cite{Hellendoorn2018}, Typilus~\cite{Allamanis2020}, and work by Xu et al.~\cite{Xu2016}.
While adding type annotations is generally considered a step forward, newly added annotations often reveal previously unnoticed type errors, which can be easily detected with a static type checker.
Unfortunately, developers commonly lack the time to fix these errors~\cite{Digrazia2022}, which hampers the usefulness of gradual typing.
%


\begin{figure}[t]
	\begin{subfigure}[t]{0.48\linewidth}
		\begin{lstlisting}[numbers=none,xleftmargin=0cm]
/*#\HLL#*/# Error: 'draw_text_rectangle' for 1st argument, expected 'str' but got 'int'./*#\HLLoff#*/
def draw_texture_rectangle(
  texture: Texture,
  scale: float = 1):
...
draw_text_rectangle(/*#\HLLL#*/scale, texture/*#\HLLLoff#*/)			
			
		\end{lstlisting}
		\caption{Code with a type error.}
		\label{fig:example1}
	\end{subfigure}
	\begin{subfigure}[t]{0.48\linewidth}
		\begin{lstlisting}[numbers=none,xleftmargin=0cm]
			
def draw_texture_rectangle(
  texture: Texture,
  scale: float = 1):
...
draw_text_rectangle(/*#\HL#*/texture, scale/*#\HLoff#*/)
		\end{lstlisting}
		\caption{Type error fixed by swapping arguments.}
		\label{fig:example2}
	\end{subfigure}
	
	\hrule
	\hrule
		
	\begin{subfigure}[t]{0.48\linewidth}
	\begin{lstlisting}[numbers=none,xleftmargin=0cm]
/*#\HLL#*/# Error: 'method_name' is declared to have type 'str' but used as type 'None'./*#\HLLoff#*/
def _decorate_async_function(
  method: Callable,
  method_name: /*#\HLLL#*/str/*#\HLLLoff#*/ = None):		
	\end{lstlisting}
	\caption{Code with a type error.}
	\label{fig:example21}
\end{subfigure}
\begin{subfigure}[t]{0.48\linewidth}
	\begin{lstlisting}[numbers=none,xleftmargin=0cm]
		
def _decorate_async_function(
  method: Callable,
  method_name: /*#\HL#*/Optional[str]/*#\HLoff#*/ = None):
	\end{lstlisting}
	\caption{Type error fixed by adding an \code{Optional} annotation.}
	\label{fig:example22}
\end{subfigure}
	\caption{Examples of type errors fixed by \name{}.}
	\label{fig:exampleintro}
\end{figure}

Figure~\ref{fig:exampleintro} shows two real-world, statically detectable type errors along with their fixes, as performed by developers.
The error presented in Figure~\ref{fig:example1} is caused by passing the arguments to a function in the wrong order~\cite{oopsla2018-DeepBugs}, i.e., a kind of problem that in statically typed languages often can be prevented by the type system.
The developers fix the problem by swapping the arguments.\footnote{\href{https://github.com/pythonarcade/arcade/commit/29972977db9e56010cd8b2e533eaa001f77114cd\#diff-0b06b8c0af34bc51343b33fd98332c989eb2322a2b3ca2afd113a14eda2cec6aL883}{https://github.com/pythonarcade/arcade/commit/c6aL883}}
The error presented in Figure~\ref{fig:example21} is caused by annotating a parameter to be a string, while at the same time, initializing it to \code{None}, which is type-incompatible with \code{str}.
To fix this error, the developer modifies the type annotation to \code{Optional[str]}.\footnote{\href{https://github.com/awslabs/aws-lambda-powertools-python/commit/5b87bb195fb154d2a112364a5d1d5c9513898e55\#diff-ac2e7c32fce52e9e17243aafe7f69a62615e44d9d3a15dabd164d16b476dc5daL523}{https://github.com/awslabs/aws-lambda-powertools-python/3898e55}}
As illustrated by these examples, there may be many ways of addressing different type errors in Python, and finding the right fix for a given error is non-trivial.

\paragraph{Significance}
Organizations with large Python code bases invest significant efforts toward using type annotations and type checkers. For example, Google's Python style guide mentions that developers are ``strongly encouraged to enable Python type analysis'' because ``The type checker will convert many runtime errors to build-time errors''.\footnote{\url{https://google.github.io/styleguide/pyguide.html\#2212-pros}}
Likewise, Dropbox type-checked over four million lines of Python in 2019, because ``A type checker will find many subtle (and not so subtle) bugs. A typical example is forgetting to handle a \emph{None} value or some other special condition''.\footnote{\url{https://dropbox.tech/application/our-journey-to-type-checking-4-million-lines-of-python}}
Finally, Meta ``use[s] it extensively to maintain the codebases of Facebook and Instagram''.\footnote{\url{https://developers.facebook.com/blog/post/2021/05/10/eli5-pyre-fast-error-flagging-python-codebases/}}

To handle type errors in legacy code and type errors revealed by adding type annotations to previously unannotated code, an automated technique to help developers fix such errors would be desirable.
However, despite the increasing popularity of automated program repair (APR)~\cite{cacm2019-program-repair}, there currently is no APR approach targeting static type errors in Python.
Compared to repair scenarios targeted by existing APR approaches, fixing type errors in Python differs in three important ways, making the problem particularly amenable to automated repair.
First, type errors require specific fix patterns, which an approach specifically targeting such errors can exploit.
Second, when a gradual type checker reports a type error, the report includes an error message that may offer hints about the location and nature of the problem.
Third, the gradual type checker also offers an automatic oracle, which an APR technique can use to validate candidate fixes.


\paragraph{Approach} This paper introduces \name{}, the first APR approach for static type errors in Python.
To guide the design of \name{}, we investigate in a preliminary study how developers typically fix type errors.
The study investigates (i) how repetitive type errors and their fixes are, (ii) how difficult it is to localize the fix location, and (iii) to what extent the error message provided by a type checker helps in finding the fix.
In short, the results show that there are recurring fix patterns, but ambiguous rules for when to apply them, and that the locations and error messages provided by the type checker are valuable information. 

Based on the results of the preliminary study, we design \name{} as a data-driven approach.
This kind of approach requires a dataset for training and evaluation. However, automatically collecting a large-scale dataset of type error fixes is challenging because it requires identifying relevant commits and isolating the type error fixes in these commits.
%
We address these challenges through an automated approach that combines gradual type checking and delta debugging~\cite{zeller2002isolating}.
Using this approach, we obtain 2,766 real-world pairs of type errors and corresponding single-hunk fixes from 176 GitHub repositories.
To the best of our knowledge, our \dataset{} dataset is the first of its kind.

The core of \name{} is a neural type error repair model.
Motivated by the relative data scarcity of the problem, we present a cross-lingual transfer learning approach.
Specifically, we base \name{} on the existing APR system TFix~\cite{Berabi2021}, which has been trained to fix linter warnings in JavaScript code.
By fine-tuning the TFix model with \dataset{}, we retain the knowledge learned from fixing JavaScript code and apply it to fixing Python type errors.
To ensure that every fix suggested by \name{} indeed fixes the targeted type error, the approach checks candidate fixes with a gradual type checker, and returns a fix only if it removes the error.

\paragraph{Results}
Our evaluation on a held-out subset of 281 type error fixes shows that \name{} finds a fix that removes type errors for 85.4\% of all errors.
Moreover, 54.4\% of the predicted fixes exactly match the developer's fix.
Comparing \name{} with previous work, we find that it clearly outperforms several state-of-the-art large language models (text-davinci-003, gpt-3.5-turbo, and gpt-4) asked to repair type errors (54.4\% vs.\ 26.4\% exact matches) and complements a technique aimed at type errors that manifest at runtime~\cite{Oh2022}.
As evidence of the usefulness of \name{} in practice, \pr{} out of 30 GitHub pull requests with \name{}-suggested fixes have been merged by the developers.
Finally, we also validate the automatically gathered \dataset{} dataset underlying our approach, and find that almost all gathered fixes are minimal and correct.

\paragraph{Contributions}
In summary, the contributions of this paper are:
\begin{itemize}
	\item An empirical study of how developers fix type errors.
	\item A technique to extract type errors and corresponding fixes through a combination of gradual type checking and delta debugging, which yields the first dataset of its kind, with 2,766 type error-fix pairs from 176 GitHub repositories.
	\item Cross-language transfer learning that uses a model pre-trained on JavaScript to repair type errors in Python.
	\item Empirical evidence of the effectiveness of the approach when being applied to real-world type errors.
\end{itemize}

	\section{Background on Python Type Checkers}
\label{sec:background}

In 2015, Python introduced a syntax for type annotations.
These annotations are optional and not checked at runtime.
The Python language also does not define a static type system, but leaves type checking to third-party tools.
In response, the Python community has developed several type checkers that perform gradual type checking~\cite{DBLP:conf/ecoop/SiekT07}, i.e., a form of type checking aimed at exposing incompatibilities between the provided type annotations while allowing parts of the program to remain unannotated.
Popular type checkers include Pyre, Mypy, Pytype, and Pyright.\footnote{\url{https://realpython.com/python-type-checking/}}
The type systems implemented by checkers differ, and hence, different type checkers may reveal different type errors~\cite{Rak-amnouykit2020}.
Conceptually, the approach described in this paper is independent of a specific type checker and could be adapted to any of the popular checkers.
Our implementation builds upon Pyre because it is widely used, available as open-source, backed by a major tech company, and has been the basis of recent work on studying Python type annotation practices~\cite{Digrazia2022}.
Pyre reports a wide range of type-related problems, such as incompatible variable, parameter, and return types, uses of unbound names, unsupported operands, and inconsistent method overrides.
We use Pyre's default configuration, i.e., it runs only on functions that are at least partially type-annotated.
In the remainder of the paper, we refer to Pyre using the term \emph{type checker}.

	\section{Preliminary Study}
\label{sec:preliminary}

To guide important design decisions of our approach, we perform a preliminary empirical study that investigates three questions (PQs):
\begin{enumerate}[leftmargin=2.25em]
	\item[PQ1] \emph{How repetitive are real-world type errors and type error fixes?}
	Answering this question is useful for deciding about the kind of technique, e.g., rule-based vs.\ data-driven, to build for automatically repairing type errors.
	\item[PQ2] \emph{How difficult is identifying the fix location for a given type error?}
	Answering this question helps us decide how \name{} can effectively determine where in the given code to fix a type error.
	\item[PQ3] \emph{How useful for fixing type errors are the error messages provided by a type checker?}
	Answering this question is useful to determine if and how a repair technique will be able to benefit from error messages.
\end{enumerate}

\subsection{Data Collection}

To address the above questions, we systematically study type error fixes in the version histories of popular projects.
We apply three strategies to select commits with type error fixes. First, we search for GitHub issues that call for help in fixing type errors. 
Second, we search for commits on GitHub via the keywords: ``type+fix'', ``pyre'' and ``mypy'' in Python repositories with more than 100 stars. 
Third, we use a dataset extracted from the top 10,000 Python repositories~\cite{Digrazia2022}, which contains commits with edits related to inserting, removing, or updating a type annotation. 

After collecting the commits, we clone the repositories and run the type checker before and after each commit.
%
During our manual inspection, we observe that some warnings and fixes are not useful for our study towards building an APR tool, and hence, we remove
(i) fixes that delete entire functions or files, without actually fixing a type error\footnote{E.g., \href{https://github.com/vkbottle/vkbottle/commit/2bc36b6d2e71e6a6d24765312cf786753201be01\#diff-c4eef9f9c1a249a379aa69f4565090841aa8dabceb6bf557c7fe469a3bc05543L21}{https://github.com/vkbottle/vkbottle/commit/2bc36b6}},
(ii) import-related warnings, as they are often due to libraries missing in the type checker's search path, and
(iii) fixes that add comments \code{\# pyre-ignore} or \code{\# type:ignore} to suppress warnings from the type checker.
Overall, for the preliminary study, we collect 125 type error fixes from 14 GitHub repositories.

\subsection{Results}

%

\subsubsection{PQ1: Repetitiveness of Type Errors and Fixes}
\label{sec:repetitiveness}

\begin{figure}[t]
	\centering
	\includegraphics[width=.99\linewidth]{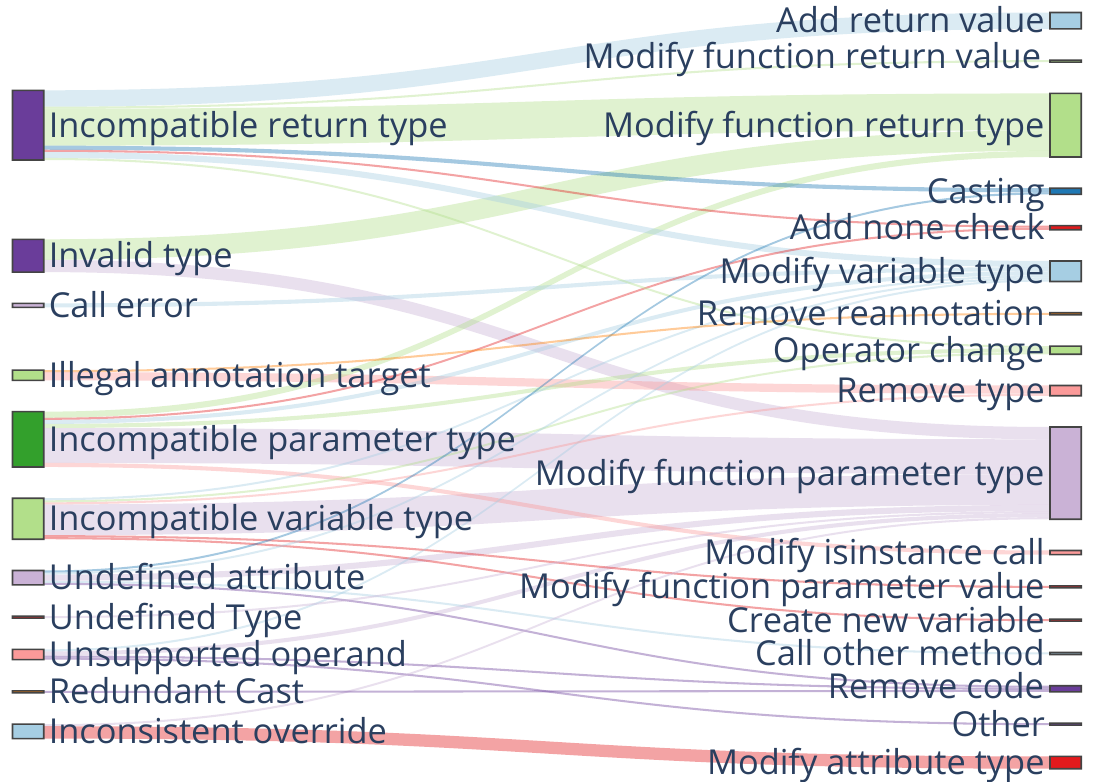}
	\caption{Type errors (left) and related fix patterns (right), based on 125 type error fixes collected in the preliminary study.}
	\label{f:type-error}
\end{figure}

We analyze the most frequent classes of type errors fixed by developers, which helps understand which errors concern developers the most, and hence, should be the focus of an APR technique. 
Figure~\ref{f:type-error} (left) shows the distribution of the most frequently fixed classes of type errors. We take the classes of type errors from the Pyre documentation.\footnote{\url{https://pyre-check.org/docs/errors/}} The most frequent classes are incompatible return, variable, and parameter types, which together account for 64.8\% of the dataset.
For example, one such fix is for a function expected to return a \code{str} but that actually returns \code{int} due to a statement \code{return -1}. The error is fixed by changing the return statement to \code{return "X"}.\footnote{\href{https://github.com/TheAlgorithms/Python/commit/97b6ca2\#diff-66f650b3a498fb126465a4b809ccb5e16f7766c633fd1f14cb06761cb880e3ccL17}{https://github.com/TheAlgorithms/Python/commit/97b6ca2}}

We also analyze the most frequent types involved in the fixes.
We observe that Python's built-in types occur frequently, e.g., \code{str} (23.9\%) and \code{int} (22.4\%).
Also relatively frequent are types related to optional values, such as \code{Optional} (9.3\%) and \code{None} (5.6\%), and other types from the \code{typing} library, e.g., \code{Union} (7.1\%).

We study how type errors are fixed by manually categorizing the error-fixing code changes into 17 classes.
This categorization was performed by one of the authors based on grounded theory~\cite{glaser1968discovery}, i.e., we discovered and refined fix patterns until they sufficiently covered the studied examples.
Figure~\ref{f:type-error} (right) shows the distribution of the identified fix patterns.
Beyond the distribution, the figure shows that there is no simple mapping from classes of type errors to fix patterns. The most frequent relationships are \emph{Incompatible return type} fixed with the pattern \emph{Modify function return type} (14.4\%) and \emph{Incompatible parameter type} fixed with the pattern \emph{Modify function parameter type} (13.6 \%).
However, the same class of type error may also get addressed by applying several other fix patterns.

\begin{finding}{PQ1}
	A few kinds of type errors account for most fixed errors, and the fixes often involve Python's built-in data types.
	The fixes expose some recurring patterns, but only an ambiguous mapping from classes of type errors to fix patterns. 
\end{finding}

\subsubsection{PQ2: Difficulty of Identifying the Fix Location}
\label{sec:fix location}

To assess the difficulty of localizing where to fix type errors, we start by investigating how much code developers typically change to fix a type error.
Based on the categorization of fix patterns in Figure~\ref{f:type-error}, we see that most fixes are single-line edits, such as modifying a type annotation from one type to another, changing an operator, removing a type annotation, or adding a cast.
%
%
Next, we study the location of the fixes (Figure~\ref{fig:preliminary}a).
More than half of the fixes happen exactly in the line where the type error is reported. Other locations include the function parameters, return annotations, and function callees (i.e., the functions that are called).

\begin{finding}{PQ2}
	Most fixes of type errors affect only a single line of code, which often is the line where the type checker reports the type error.
\end{finding}

\subsubsection{PQ3: Usefulness of Error Messages and Locations}

\label{sec:preliminarpyre}

\begin{figure*}[t]
	\begin{subfigure}[t]{0.48\linewidth}
		\begin{lstlisting}[numbers=none,xleftmargin=0cm]
def is_valid_public_key_static(
  local_private_key_str: str, /*#\HLLL#*/remote_public_key_str: str/*#\HLLLoff#*/, prime: int
) -> bool:
  ...
/*#\HLL#*/[Error message] Expected `int` for 1st parameter but got `str`./*#\HLLoff#*/   
  if pow(remote_public_key_str, (prime - 1) // 2, prime) == 1:
  ...
		\end{lstlisting}
		\caption{Commit with a type error.}
		\label{fig:example14}
	\end{subfigure}
	\textcolor{lightgray}{\vrule} \
	\begin{subfigure}[t]{0.48\linewidth}
		\begin{lstlisting}[numbers=none,xleftmargin=0cm]
def is_valid_public_key_static(
  /*#\HL#*/remote_public_key_str: int/*#\HLoff#*/, prime: int
) -> bool:
  ...
/*#\HLL#*/[Fix pattern] Modify function parameter type./*#\HLLoff#*/
  if pow(remote_public_key_str, (prime - 1) // 2, prime) == 1:
  ...
		\end{lstlisting}
		\caption{Commit that applies the ``Use expected type'' pattern.}
		\label{fig:example15}
	\end{subfigure}
	
	\hrule
	
	\begin{subfigure}[t]{0.48\linewidth}
	\begin{lstlisting}[numbers=none,xleftmargin=0cm]
def get_model_for_finetuning(
  previous_model_file: Optional[Union[Path, Text]]
/*#\HLL#*/[Error message] Expected 'Optional[Text]', got 'Union[None, Path, Text]'/*#\HLLoff#*/
) -> Optional[/*#\HLLL#*/Text/*#\HLLLoff#*/]:
	\end{lstlisting}
	\caption{Commit with a type error.}
	\label{fig:example16}
\end{subfigure}
\textcolor{lightgray}{\vrule} \
\begin{subfigure}[t]{0.48\linewidth}
	\begin{lstlisting}[numbers=none,xleftmargin=0cm]
def get_model_for_finetuning(
  previous_model_file: Optional[Union[Path, Text]]
/*#\HLL#*/[Fix pattern] Modify function return type./*#\HLLoff#*/
) -> Optional[/*#\HL#*/Union[Path, Text]/*#\HLoff#*/]:
	\end{lstlisting}
	\caption{Commit that applies the ``Do not use expected type'' pattern.}
	\label{fig:example17}
\end{subfigure}
	\caption{Examples of fixing type errors based on error messages.}
	\label{fig:examplepyre}
\end{figure*}

Finally, we want to understand how useful the error messages provided by a type checker are for fixing type errors. To this end, we extract from the error message the kind of error, the types involved, and any hints about the location and the fix.
We classify an error message as \emph{correctly hinted} if the message contains the type that the developer uses to fix the error. Figure~\ref{fig:example14} shows an example, where the type checker returns the following error message: ``Incompatible parameter type [6]: Expected \code{int} for 1st positional only parameter to call \code{pow} but got \code{str}'', where the message hints at replacing \code{str} with the correct type \code{int}.\footnote{\href{https://github.com/TheAlgorithms/Python/commit/60895366c0f50844af2737130ed98c2510e90060\#diff-b146f69681cf682252a7ed83539b858a4a1b2c9367dcbc172f064637c26d4eceL244}{https://github.com/TheAlgorithms/Python/commit/6089536}} Note that the hinted type might not be an exact match to the newly annotated type. For example, we consider an error message that suggests \code{str} as correctly hinted also if the developer fixes the error using \code{Optional[str]}.

\begin{figure*}
	\captionsetup[subfigure]{justification=centering}
	\centering
	\includegraphics[width=1\linewidth]{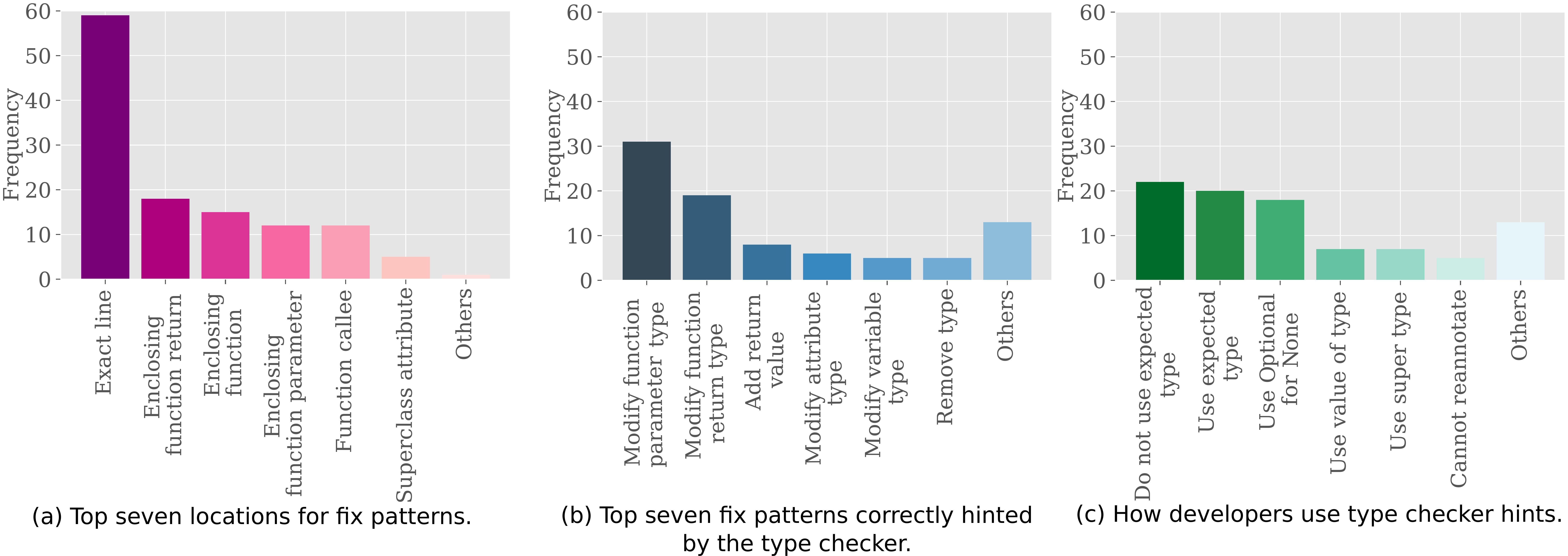}
	\caption{Fix locations and usefulness of error messages.}
\label{fig:preliminary}
\end{figure*}

Given the above definition, we find that 89 out of 125 (71.2\%) type fixes in our study are correctly hinted by the type checker.
These hinted types can serve as a reference for APR tools to narrow down the search space. The types of code changes correctly hinted by the checker are shown in Figure~\ref{fig:preliminary}b.

	

Figure~\ref{fig:preliminary}c shows how the developers use the correctly hinted types. 
As an example, in Figure~\ref{fig:example16}, the type checker returns the following error message: ``Incompatible return type [7]: Expected \code{Optional[Text]} but got \code{Union[None, Path, Text]}''.\footnote{\href{https://github.com/RasaHQ/rasa/commit/1ded5ef\#diff-748793f55fcf9a17f1c97bf1b4e2d48bd8c3629768d816dbec5328690733d0beL624}{https://github.com/RasaHQ/rasa/commit/1ded5ef}}
The developer does not fix the error by using the suggested type \code{Union[None, Path, Text]}, but instead uses \code{Optional[Union[Path, Text]]}. In contrast, the example in Figure~\ref{fig:example15} shows a case where the developer uses the type suggested by the type checker.
We find that for 64 out of the 89 hints (71.9\%) the type used by the developer is exactly as suggested in the error message. It is also common to introduce a value of the suggested type, e.g., by adding \code{return -1} to a function supposed to return \code{int}. 
%
Besides the 89 error messages that correctly hint at the correct type, most of the remaining messages (24 out of 36) give no hint at all.
For example, this is the case for the error classes ``Undefined type'', ``Invalid type'', and ``Undefined attribute''.


\begin{finding}{PQ3}
	Most types used in fixes (71.2\%) are correctly hinted by the type checker, and developers often follow these hints.
\end{finding}

\subsection{Implications}

The three main findings of the preliminary study guide the design of our approach as follows.
(PQ1)~We observe that type errors and their fixes expose recurring patterns, which might suggest an approach based on manually designed rules and heuristics for selecting them.
However, we also find that there is only an ambiguous mapping from errors to fix patterns, making a rule-based approach laborious and fragile.
As a result, we decide against a rule-based and in favor of a data-driven approach, aiming for a model that learns when to apply which fix pattern from fixes performed by developers.
(PQ2)~We find that most type errors are fixed by editing a single line, and that this line is often localized correctly by the type checker.
Hence, we focus our work on fixing type errors in single-hunk edits\footnote{Hunks may be larger than single lines, allowing \name{} to predict some fixes that involve multiple lines.} and exploit the localization hint given by the type error location.
(PQ3)~We find that the error message provided by the type checker often gives valuable hints for finding the fix, e.g., which type to use.
As a result, we provide the error message as an input to our approach.

	\section{Approach}
\label{sec:datasetextracted}

\begin{figure}[t]
	\centering
	\includegraphics[width=.9\linewidth]{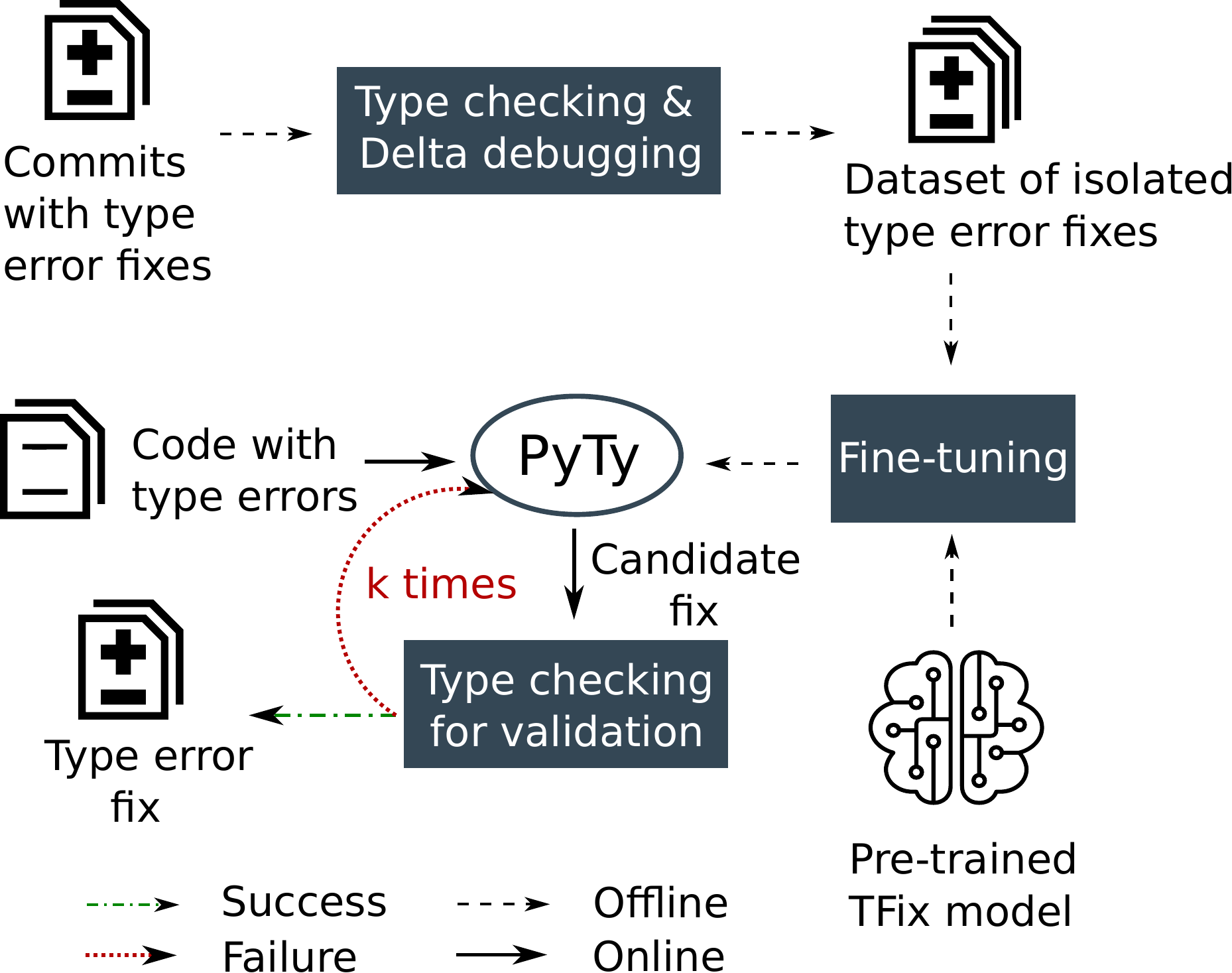}
	\caption{Overview of the approach.}
	\label{fig:overview}
\end{figure}

Based on the findings of our preliminary study, we design \name{}, a data-driven approach to automatically fix static type errors in Python using a cross-language transfer learning approach. 
Figure~\ref{fig:overview} shows an overview of the approach, which consists of two phases.
First, during the offline phase, we automatically collect a dataset of type error fixes from GitHub, which we call \dataset{}, by combining delta debugging and gradual type checking, followed by fine-tuning a pre-trained model~\cite{Berabi2021} with \dataset{}.
Second, during the online phase, \name{} receives code with a type error as the input and then queries the model for fix candidates.
The approach uses the type checker to validate that the type error gets resolved when applying a fix candidate, and then reports only fixes that are guaranteed to remove the targeted type error.

\subsection{Automated Data Gathering}
\label{sec:delatadebugging}
To build a learning-based APR model, we must first collect a relevant dataset as our training data.
As a first step, we search for Python repositories that are popular ($\ge100$ stars), have a manageable size ($\leq5$GB), and were created between 2010 and 2021 on GitHub.
We use the keywords ``fixing+typing'', ``fixing+pyre'', ``fixing+mypy'', ``typing+bug'', and ``typing+error'' to search for commits that possibly remove type errors.
Next, we run the type checker before and after each such commit to find commits that indeed remove type errors. 
As a result, we obtain 32,330 type errors that are removed by 4,515 commits in 176 GitHub repositories.


Many of the extracted commits contain changes not directly related to fixing type errors.
Moreover, a single commit often fixes multiple type errors.
Figure~\ref{fig:exampledeltadeb} illustrates these problems with an example.\footnote{Simplified from \href{https://github.com/jazzband/django-redis/commit/5f6f38362dd587aae78d9b8ff97a1e2fe800ba5d\#diff-680e70773dd1969f27b4d66da3dd5759928346dceb2599182141ecbc7894764cL17}{https://github.com/jazzband/django-redis/commit/5f6f383}}
To isolate individual type error fixes, we present a delta debugging-inspired~\cite{zeller2002isolating} algorithm that reduces commits into small code changes that fix a single type error.
The basic idea is to iteratively reduce the set of code hunks while preserving the fact that the code change fixes a particular type error.


\begin{algorithm}[tb]
	\small
	\begin{algorithmic}[1]	
		\Require{Files $f_{\mathit{old}}$ with type error $err$ and $f_{\mathit{new}}$ after commit with $err$ fixed}
		\Ensure{Minimal hunk(s) of the commit, containing only $err$ fixed}
		\State $W \leftarrow \mathit{type\_check}(f_{\mathit{old}})$    \Comment{Set of all warnings in $f_{\mathit{old}}$}
		\State $D_{\mathit{original}} \leftarrow \mathit{diff}(f_{\mathit{old}},f_{\mathit{new}})$  \Comment{All diff hunks between $f_{\mathit{old}}$ and $f_{\mathit{new}}$}
		\State $\mathit{granularity} \leftarrow 2$ \Comment{Set default granularity}
		\While{$\mathit{granularity}\le \mathit{size}(D_{\mathit{original}})$}
		\State $\mathit{min} \leftarrow \mathit{False}$		
		\State $D \leftarrow D_{\mathit{original}}$
		\While{$\mathit{size}(D) > 1$ \mbox{\textbf{and}} $\mathit{granularity} > 1$}
		\For{$d$ \mbox{\textbf{in}} $\mathit{split}(D,\mathit{granularity})$}\Comment{Split the set of hunks $D$}
		\State $f_\mathit{fixed} \leftarrow \mathit{patch}(f_{\mathit{old}},d)$ \Comment{Apply a subset of $D$ to file $f_{\mathit{old}}$}
		\If{$\mathit{parsable}(f_\mathit{fixed})$}
		\State $W_\mathit{fixed} \leftarrow \mathit{type\_check}(f_\mathit{fixed})$
		\If{$\not\exists~err$ \mbox{\textbf{in}} $W_\mathit{fixed}$ \mbox{\textbf{and}} $W_\mathit{fixed} == W$}
		\If{$\mathit{size}(d) == 1$}
		\State \Return $d$
		\Else
		\State $D \leftarrow d$
		\State $\mathit{min} \leftarrow \mathit{True}$
		\State \mbox{\textbf{break}}
		\EndIf
		\EndIf
		\EndIf
		\EndFor	
		\If{$\mathit{min} == \mathit{False}$}
		\If{$\mathit{granularity} * 2 \le \mathit{size}(D)$}
		\State $\mathit{granularity} \leftarrow \mathit{granularity} * 2$
		\ElsIf{$\mathit{granularity} == \mathit{size}(D)$}
		\State \Return D
		\Else 
		\State  $\mathit{granularity} = \mathit{size}(D)$
		\EndIf
		\EndIf
		\EndWhile
		\State $f_\mathit{fixed} \leftarrow \mathit{patch}(f_{\mathit{old}},D)$ \Comment{Apply $D$ (size=1) to file $f_{\mathit{old}}$}
		\If{$\mathit{parsable}(f_\mathit{fixed})$}
		\State $W_\mathit{fixed} \leftarrow \mathit{type\_check}(f_\mathit{fixed})$
		\If{$\not\exists~e$ \mbox{\textbf{in}} $W_\mathit{fixed}$ \mbox{\textbf{and}} $W_\mathit{fixed} - W$ = $\emptyset$}
		\State \Return $D$
		\EndIf
		\EndIf
		\EndWhile
	\end{algorithmic}
	\caption{Extract relevant hunks with delta debugging.}\label{alg:cap}
\end{algorithm}

Algorithm~\ref{alg:cap} summarizes our approach for reducing a commit to a small set of code changes that fix the given type error.
We illustrate the algorithm using the example in Figure~\ref{fig:exampledeltadeb}.
We focus on the type error ``Unbound name: \code{basestring} is used but not defined in the current scope'', reported for the line in hunk~H1.
The error gets fixed by changing the base class to \code{object}.
Our approach considers the four code hunks (H1, H2, H3, and H4) of this commit, and determines which hunks are relevant for fixing the type error with the following steps, where line numbers refer to Algorithm~\ref{alg:cap}:
\begin{enumerate}
	\item The algorithm splits the set of hunks into \code{H1+H2} and \code{H3+H4} with granularity two (line~8):
	\begin{enumerate}
		\item The algorithm patches the code with only hunks \code{H1+H2}, which yields parsable code (lines~9 and~10).
		\item The type error disappears and there are no new errors (lines~11 and~12).
		\item There are still two hunks \code{H1+H2} (line~13).
	\end{enumerate}
	\item The algorithm splits the code hunks \code{H1+H2} into \code{H1} and \code{H2} (line~8): \begin{enumerate}
		\item The algorithm patches the code with only hunk \code{H1}, which yields parsable code (lines~9 and~10).
		\item The type error disappears and there are no new errors (lines~11 and~12).
	\end{enumerate}
	\item The algorithm returns \code{H1} as a minimal code change to fix the type error (line~14).
\end{enumerate}

\begin{figure}[]
	\begin{subfigure}[t]{0.483\linewidth}
		\begin{lstlisting}[numbers=none,xleftmargin=0cm, basicstyle=\ttfamily\footnotesize, keywordstyle=\ttfamily\footnotesize]
# Hunk H1
class CacheKey(/*#\HLLL#*/basestring/*#\HLLLoff#*/):

# Hunk H2
  /*#\HLLL#*/pass/*#\HLLLoff#*/
  
  
  
# Hunk H3
  if isinstance(key, CacheKey):
    key = CacheKey(/*#\HLLL#*/smart_str(key)/*#\HLLLoff#*/)
    
# Hunk H4
  if timeout /*#\HLLL#*/== 0:
/*#\HLLLoff#*/
		\end{lstlisting}
	\caption{Commit 1.}
		\label{fig:example3}
	\end{subfigure}
	\textcolor{lightgray}{\vrule} \
	\begin{subfigure}[t]{0.42\linewidth}
		\begin{lstlisting}[numbers=none,xleftmargin=0cm, basicstyle=\ttfamily\footnotesize, keywordstyle=\ttfamily\footnotesize]
# Hunk H1
class CacheKey(/*#\HL#*/object/*#\HLoff#*/):

# Hunk H2
  /*#\HL#*/def __init__(self, key):
    self._key = key
    .../*#\HLoff#*/

# Hunk H3
  if /*#\HL#*/not/*#\HLoff#*/ isinstance(key, CacheKey):
    key = CacheKey(/*#\HL#*/key/*#\HLoff#*/)

# Hunk H4
  if timeout /*#\HL#*/is None:
    .../*#\HLoff#*/
		\end{lstlisting}
	\caption{Commit 2.}
		\label{fig:example4}
	\end{subfigure}
	\caption{Multi-hunk commit that fixes multiple type errors.}
	\label{fig:exampledeltadeb}
\end{figure}

To properly track the error location while reducing the hunks, we need to keep track of how the line numbers change. To this end, we calculate the new line number based on how many lines are inserted or removed in each code hunk. We consider the error fixed if the error no longer exists at the corresponding line and column. If the error is located inside a code hunk, i.e., the code with the error is being modified, we consider the error as fixed only if all lines in the code hunk are free of errors after the change.

To ensure the quality of \dataset{}, we apply additional filtering steps.
Algorithm~\ref{alg:cap} checks that there are no new errors introduced by the code changes (line 12).
The algorithm also rejects any set of code hunks that result in parsing failures (line 10). 
%
The space complexity of the algorithm is $\mathcal{O}(2*N)$, where $N=size(D_{original})$ and the time complexity is $\mathcal{O}(N*\log{}N)$.

Running the algorithm on the 32,330 type errors gives 11,955 examples of reduced error fixes. We further filter them by keeping only fixes that (i) are relatively small ($\leq$ 512 characters and at most three changed lines), which is motivated by limitations of the neural model (Section~\ref{sec:model}); (ii) do not contain any error suppression; (iii) are not only deletion; (iv) are located close (i.e., within the same hunk) to the reported bug location.
Finally, after applying these filters, \dataset{} has 2,766 entries that cover ten frequent categories of type errors listed in Table~\ref{table:resulttop1}.\footnote{The distribution of type errors is similar to that in our preliminary study (Figure~\ref{f:type-error}), but not exactly the same because the datasets differ.}
We select 10\% (always rounding up to the nearest integer) of the entries of each error class in this final dataset as a test set, and then split the remaining fixes into 90\% for training and 10\% for validation. 


\subsection{Neural Type Error Fixing}
\label{sec:model}

Given the automatically extracted dataset of type error fixes, \name{} trains a neural model that predicts how to fix type errors.
The input to the model is a sequence of one or more lines, i.e., the size of a single hunk, that contains a type error.
The output of the approach is a fix that removes the targeted type error.


\subsubsection{Base Model}
Instead of learning a model from scratch, we fine-tune a model pre-trained on another APR task.
Building on a pre-trained model is motivated by the fact that \dataset{}, with 2,766 examples, is relatively small.
As our base model, we use TFix~\cite{Berabi2021}, a learning-based APR technique trained to fix linter errors in JavaScript.
We select TFix for three reasons: (i) it is already trained on a bug fix dataset of 104,804 samples, (ii) it accepts error messages as input, and (iii) the TFix authors used it to predict single line fixes, which resembles our single-hunk setup.
By fine-tuning TFix, \name{} transfers the already learned knowledge to the related but different domain of Python type errors (Section~\ref{sec:evalablation}).
TFix itself is based on T5~\cite{Raffel2020}, a transformer-based model that maps sequences of input tokens to sequences of output tokens.
The simple input and output structure eliminates the need for implementing a static analysis tool to transform our code into a specific structure, such as graphs~\cite{dinella2020hoppity,allamanis2017learning}.
Furthermore, having an unconstrained token sequence may enable the model to fix errors missed by a template-based APR approach, which is inherently limited to its set of templates (Section~\ref{sec:eval pyter}).
 
\subsubsection{Fine-Tuning for Python Type Error Fixing}
To fine-tune TFix with \dataset{}, we follow the same input format as TFix:\\
$\text{``fix''} \vartextvisiblespace t \vartextvisiblespace m \vartextvisiblespace l_k \vartextvisiblespace \text{``:''} \vartextvisiblespace C$, where ``fix'' and ``:'' are literals, $t$ is the class of type error, $m$ is the error message, $l_k$ is the line of code with the type error,  $\vartextvisiblespace$ represents a space, and $C$ represents the buggy lines of code (i.e., the single-hunk we extracted in Section~\ref{sec:delatadebugging}). In the T5 framework, the string `$\text{``fix''} \vartextvisiblespace t \vartextvisiblespace m \vartextvisiblespace l_k \vartextvisiblespace \text{``:''}$` represents the current task, and $C$ represents the input of this task. The model outputs $C'$, which we use as a replacement for $C$ to fix the type error.


\subsubsection{Python Code Pre- and Post-Processing}
\label{sec:preprocess}
We use the tokenizer from the Python standard library to pre-process the source code and inject special tokens for indentation and dedentation.
TFix uses SentencePiece~\cite{kudo2018sentencepiece} as its tokenizer.
However, SentencePiece does not take the number of white\-spaces into account, as it escapes all white\-spaces into a single ``\_'' symbol.
Since the amount of whitespace carries semantics in Python, we preserve this information by adding special tokens ``\textless IND\textgreater'' and ``\textless DED\textgreater'' into the source code before passing it to the model.
Given a prediction by the model, \name{} replaces the special tokens in a post-processing step to obtain syntactically correct Python code.


\subsubsection{Validating Fixes via Type Checking}

Once trained, we query the model for a ranked list of the $k$ most likely fixes.
To ensure that a fix suggestion given to a user indeed removes the targeted type error, \name{} validates all candidate fixes by running the type checker on them.
If and only if the targeted type error disappears and no new errors appear, the fix is suggested to the user.


\section{Implementation} 
\label{sec:implementation}
We fine-tune the t5-base (220M parameters) model of TFix for 30 epochs with a batch size of 32, and then evaluate the model that has the lowest validation loss on the validation set. The model converges at the 17th epoch. We follow the default hyperparameters of TFix~\cite{Berabi2021}. 
When validating candidate fixes using the type checker, we sample from the model up to $k=50$ predictions to be validated.
Since \name{} validates fix candidates automatically, a user does not have to inspect these 50 suggestions, but only the first one found to successfully remove the type error.
To fix 281 type errors (i.e., our test set, which amounts to a total of 174,586 lines of code) and automatically check whether the targeted type errors disappear, \name{} takes in total six hours and 44 minutes, i.e., an average of 86.2 seconds per type error fix.
We perform all experiments on a server with 48 Intel Xeon CPU cores clocked at 2.2GHz, 250GB of RAM, one NVIDIA Tesla V100 GPU, running Ubuntu 18.04.
Most of the time is spent on running the type checker for validating candidate fixes. 

	\section{Evaluation}
We evaluate our \dataset{} dataset and \name{}, our learning-based type error repair approach, focusing on the following research questions (RQs):
\begin{itemize}[leftmargin=2.25em]
	\item[RQ1] How effective is our automated data gathering at producing minimal code changes that fix type errors?
	\item[RQ2] How effective is \name{} at fixing type errors?
	\item[RQ3] How do variants of \name{} compare to the full approach?
	\item[RQ4] How does \name{} compare to state-of-the-art APR techniques?
\end{itemize}

\subsection{RQ1: Effectiveness of Automatic Data Gathering}
\subsubsection{Data analysis}
To validate the effectiveness of automatically gathering \dataset{}, two of the authors independently annotate a random sample of 100 of the 2,766 entries in the dataset.
The sample contains at least one error from each class of type errors except for ``call error''.
Each entry is assigned one of three labels:
\emph{minimal} if the extracted code change fixes a type error and cannot be further reduced,
\emph{correct but not minimal} if the extracted code change correctly fixes a type error but is not minimal, and \emph{wrong} otherwise.


\subsubsection{Results} After independently labeling the 100 entries, the two annotators initially agree on 89 labels. 
After discussing the divergent labels and refining the labels of some entries, there is a final agreement on 94/100 \emph{minimal},  3/100 \emph{correct but not minimal}, and 0/100 \emph{wrong} entries. 
The remaining three entries with divergent labels are due to hunks that fix two type errors at once.
These entries are minimal in the sense that a hunk-based reduction algorithm cannot further reduce them, but they could be further reduced by a more fine-grained reduction algorithm~\cite{ase2017-GTR,DBLP:conf/icse/SunLZGS18}.
The inter-rater agreement, as given by Cohen's kappa coefficient~\cite{cohen1960coefficient} is 0.651, which means a \emph{substantial agreement}~\cite{Landis1977}.

As an example of a \emph{minimal} type error fix, recall hunk \code{H1} from the previously discussed commit in Figure~\ref{fig:exampledeltadeb}.
All changes in hunk H1 are necessary for fixing the type error.
Figure~\ref{fig:manualeval2} shows an example of a \emph{correct but not minimal} reduced commit, which includes some changes not relevant to fixing the type error.\footnote{\href{https://github.com/kinnala/scikit-fem/commit/961610a6dc10fd23cb777540eeb0cc2cda555ca3}{https://github.com/kinnala/scikit-fem/commit/a555ca3}}
A single hunk updates multiple parameter type annotations of the same function.
However, only one code change is relevant to fixing the type error reported for \code{basis\_to}, which should be annotated \code{Optional[Basis]} instead of \code{Basis}, as it is initialized to \code{None}.

\begin{figure}[]
	\begin{subfigure}[t]{0.48\linewidth}
		\begin{lstlisting}[numbers=none,xleftmargin=0cm]
basis_from: /*#\HLLL#*/Basis/*#\HLLLoff#*/ = None,
basis_to: /*#\HLLL#*/Basis/*#\HLLLoff#*/ = None,
I: /*#\HLLL#*/ndarray/*#\HLLLoff#*/ = None,
expand: bool = False) -> ndarray:
		\end{lstlisting}
		\caption{Commit with multiple type errors.}
		\label{fig:example12}
	\end{subfigure}
	\textcolor{lightgray}{\vrule} \
	\begin{subfigure}[t]{0.48\linewidth}
		\begin{lstlisting}[numbers=none,xleftmargin=0cm]
basis_from: /*#\HL#*/Optional[Basis]/*#\HLoff#*/ = None,
basis_to: /*#\HL#*/Optional[Basis]/*#\HLoff#*/ = None,
I: /*#\HL#*/Optional[ndarray]/*#\HLoff#*/ = None,
expand: bool = False) -> ndarray:
		\end{lstlisting}
		\caption{Commit with multiple type error fixes.}
		\label{fig:example13}
	\end{subfigure}
	\caption{Example of a correct (but not minimal) entry in \dataset{}.}
	\label{fig:manualeval2}
\end{figure}

\begin{figure*}[t]
	\begin{subfigure}[t]{0.33\linewidth}
		\begin{lstlisting}[numbers=none,xleftmargin=0cm,breaklines=true]
vprint(f"{prefix} {lineno}: {action_name} 
  Constrain Mouse: {'yes' if constraint > 0 
  else ('no' if /*#\HLLL#*/constrained/*#\HLLLoff#*/ == 0 else 'check stack')}")
		\end{lstlisting}
		\caption{Code with type error.}
		\label{fig:exactmatch1}
	\end{subfigure}
	\textcolor{lightgray}{\vrule} \
	\begin{subfigure}[t]{0.32\linewidth}
		\begin{lstlisting}[numbers=none,xleftmargin=0cm,breaklines=true]
vprint(f"{prefix} {lineno}: {action_name} 
  Constrain Mouse: {'yes' if constraint > 0 
  else ('no' if /*#\HL#*/constraint/*#\HLoff#*/ == 0 else 'check stack')}")
		\end{lstlisting}
		\caption{Fix by the developer.}
		\label{fig:exactmatch2}
	\end{subfigure}
	\textcolor{lightgray}{\vrule} \
	\begin{subfigure}[t]{0.32\linewidth}
		\begin{lstlisting}[numbers=none,xleftmargin=0cm,breaklines=true]
vprint(f"{prefix} {lineno}: {action_name} 
  Constrain Mouse: {'yes' if constraint > 0 
  else ('no' if /*#\HL#*/constraint/*#\HLoff#*/ == 0 else 'check stack')}")
		\end{lstlisting}
		\caption{Fix suggested by \name{}.}
		\label{fig:exactmatch3}
	\end{subfigure}
	\caption{Exact match of fix for type error ``Unbound name: Name \code{constrained} is used but not defined in the current scope''.}
	\label{fig:exactmatch}
\end{figure*}
\begin{figure*}[t]
	\begin{subfigure}[t]{0.32\linewidth}
		\begin{lstlisting}[numbers=none,xleftmargin=0cm,breaklines=true]
...
/*#\HLLL#*/string = _fmt(string)/*#\HLLLoff#*/
return lib.TCOD_console_get_height_rect_fmt(
  self.console_c, x, y, width, height, /*#\HLLL#*/string/*#\HLLLoff#*/)
		\end{lstlisting}
		\caption{Code with type error.}
		\label{fig:naturalfix1}
	\end{subfigure}
	\textcolor{lightgray}{\vrule} \
	\begin{subfigure}[t]{0.32\linewidth}
		\begin{lstlisting}[numbers=none,xleftmargin=0cm,breaklines=true]
...

return lib.TCOD_console_get_height_rect_fmt(
  self.console_c, x, y, width, height, /*#\HL#*/_fmt(string)/*#\HLoff#*/)
		\end{lstlisting}
		\caption{Fix by the developer.}
		\label{fig:naturalfix2}
	\end{subfigure}
	\textcolor{lightgray}{\vrule} \
	\begin{subfigure}[t]{0.32\linewidth}
		\begin{lstlisting}[numbers=none,xleftmargin=0cm,breaklines=true]
...
/*#\HL#*/byte_string = _fmt(string)/*#\HLoff#*/
return lib.TCOD_console_get_height_rect_fmt(
  self.console_c, x, y, width, height, /*#\HL#*/byte_string/*#\HLoff#*/)
		\end{lstlisting}
		\caption{Fix suggested by \name{}.}
		\label{fig:naturalfix3}
	\end{subfigure}
	\caption{Correct fix different from the developer-provided fix for type error ``Incompatible variable type: \code{string} is declared to have type \code{str} but is used as type \code{bytes}''.}
	\label{fig:naturalfix}
\end{figure*}

\begin{figure*}[t]
	\begin{subfigure}[t]{0.32\linewidth}
		\begin{lstlisting}[numbers=none,xleftmargin=0cm,breaklines=true]
},
/*#\HLLL#*/F5_DEVICE_TYPE:/*#\HLLLoff#*/ { 
  DEVICE_CLASS_KEY: F5Device,
		\end{lstlisting}
		\caption{Code with type error.}
		\label{fig:cannotfix1}
	\end{subfigure}
	\textcolor{lightgray}{\vrule} \
	\begin{subfigure}[t]{0.32\linewidth}
		\begin{lstlisting}[numbers=none,xleftmargin=0cm,breaklines=true]
},
/*#\HL#*/F5_API_DEVICE_TYPE:/*#\HLoff#*/ { 
	DEVICE_CLASS_KEY: F5Device,
		\end{lstlisting}
		\caption{Fix by the developer.}
		\label{fig:cannotfix2}
	\end{subfigure}
	\textcolor{lightgray}{\vrule} \
	\begin{subfigure}[t]{0.32\linewidth}
		\begin{lstlisting}[numbers=none,xleftmargin=0cm,breaklines=true]
},
/*#\HL#*/DEVICE_TYPE:/*#\HLoff#*/ { 
	DEVICE_CLASS_KEY: F5Device,
	\end{lstlisting}
	\caption{Fix suggested by \name{}.}
	\label{fig:cannotfix3}
\end{subfigure}
\caption{Fix predicted by the neural model, but not suggested to the user, as the type error ``Unbound name: Name \code{F5\_DEVICE\_TYPE} is used but not defined in the current scope'' would still exist for \code{DEVICE\_TYPE}.}
\label{fig:cannotfix}
\end{figure*}

\begin{figure*}[t]
	\begin{subfigure}[t]{0.32\linewidth}
		\begin{lstlisting}[numbers=none,xleftmargin=0cm,breaklines=true]
global Bot
if self is Bot:
 
  Bot = new
		\end{lstlisting}
		\caption{Code with type error.}
		\label{fig:unnaturalfix1}
	\end{subfigure}
	\textcolor{lightgray}{\vrule} \
	\begin{subfigure}[t]{0.32\linewidth}
		\begin{lstlisting}[numbers=none,xleftmargin=0cm,breaklines=true]
global Bot
if self is Bot:
  /*#\HL#*/assert isinstance(new, BotUser)/*#\HLoff#*/
  Bot = new 
		\end{lstlisting}
		\caption{Fix by the developer.}
		\label{fig:unnaturalfix2}
	\end{subfigure}
	\textcolor{lightgray}{\vrule} \
	\begin{subfigure}[t]{0.32\linewidth}
		\begin{lstlisting}[numbers=none,xleftmargin=0cm,breaklines=true]
global Bot
if self is Bot:
 
  /*#\HL#*/new_Bot = new/*#\HLoff#*/
		\end{lstlisting}
		\caption{Fix suggested by \name{}.}
		\label{fig:unnaturalfix3}
	\end{subfigure}
	\caption{\name{}-suggested fix that removes the error ``Incompatible variable type: \code{Bot} is declared to have type \code{BotUser} but is used as type \code{User}'', while changing the behavior in an unintended way.}
	\label{fig:unnaturalfix}
\end{figure*}

\begin{table}[t]
	\centering
	\caption{Results of \name{} for each class of type error.}
	\label{table:resulttop1}
    \setlength{\tabcolsep}{1.3pt}
	\begin{tabular}{@{}lrrr@{}}
		\toprule
		         &  Samples\phantom{---} & \multicolumn{2}{c}{Effectiveness of \name{}}                            \\
		         \cmidrule{3-4}
		Classes of type errors                                       & (test set)\phantom{---} & Error & Exact \\
		&&removal&match\\
		\midrule
		Incompatible variable type                                                   & 821\phantom{---}(83)              & 90.4\%                   & 65.1\%                     \\
		Incompatible parameter type                                                  & 600\phantom{---}(60)              & 80.0\%                  & 36.7\%                     \\
		Incompatible return type                                                     & 296\phantom{---}(30)              & 73.3\%                   & 43.3\%                     \\
		Invalid type                                                                    & 291\phantom{---}(30)              & 100.0\%                   & 83.3\%                     \\
		Unbound name                                                                    & 258\phantom{---}(26)             & 76.9\%                   & 42.3\%                     \\
		Incompatible attribute type                                                    & 258\phantom{---}(26)             & 92.3\%                   & 73.1\%                     \\
		Unsupported operand                                                              & 124\phantom{---}(13)             & 76.9\%                   & 38.5\%                    \\
		Strengthened precondition                            & 59\phantom{-----}(6)             & 83.3\%                    & 50.0\%                    \\
		Weakened postcondition                                & 51\phantom{-----}(6)              & 50.0\%                   & 0.0\%                     \\
		Call error                                                                     & 8\phantom{-----}(1)                & 100.0\%                  & 100.0\%                    \\ \midrule
		Total                                                                      & \textbf{2,766 (281)}   & \textbf{85.4\%}          & \textbf{54.4\%}            \\ \bottomrule
	\end{tabular}
\end{table}

\begin{finding}{RQ1}
The automated data gathering yields type error fixes that are mostly correct (97/100) and minimal (94/100), i.e., \dataset{} provides a solid basis to train and validate \name{}.
\end{finding}

\subsection{RQ2: Effectiveness of \name{}}
\label{sec:rq2}
We evaluate the effectiveness of \name{} on all ten classes of type errors covered by our test set.
We configure the approach to consider up to $k=50$ candidate fixes.
Note that users do not have to manually check all candidate fixes, but only see the first successful fix.

\subsubsection{Metrics}
We use two metrics to evaluate the effectiveness of \name{}.
First, we compute the \emph{error removal rate}, i.e., how often the approach succeeds at finding a fix that removes the targeted type error without introducing new type errors.
Second, we compute the \emph{exact match rate}, i.e., how often the model output is identical to the fix committed by the developer.
This metric underapproximates the abilities of \name{}, as there might be fixes that address the type error in a reasonable way that differs from the original fix. 

\subsubsection{Quantitative Results}
Table~\ref{table:resulttop1} shows the number of samples used for training and testing, the error removal rate, and the exact match accuracy.
Each row in the table corresponds to one kind of type error reported by the type checker.
\name{} successfully removes the type error in 85.4\% of the cases, and it 
finds exactly the developer-provided fix for 54.4\% of all errors.
Comparing different kinds of type errors, we find the approach to be effective across a wide range of errors.
An exception are \emph{Weakened postcondition} errors, which are often caused by type-incorrect, overriding methods in custom classes, i.e., a kind of mistake that requires non-local, project-specific information to be fixed.

\subsubsection{Examples}
\label{sec:cases}
We illustrate the strengths and limitations of \name{} with four representative examples.
Figure~\ref{fig:exactmatch} shows an exact match of the developer fix.\footnote{\href{https://www.github.com/DragonMinded/bemaniutils/commit/72f81e4f58c2ef9d2b51f63ed7b52fbd9438a3da}{https://www.github.com/DragonMinded/bemaniutils/commit/438a3da}}
The error is because the variable \code{constrained} used in the format string is not defined. \name{} successfully fixes the mistake by replacing \code{constrained} with \code{constraint}, which exactly matches the developer’s fix.

The example in Figure~\ref{fig:naturalfix} fixes the type error in a way that matches the intention of the developer but differs from the original fix.\footnote{\href{https://github.com/libtcod/python-tcod/commit/60066f30f07303a0cb7092b760a8e661330a63b9}{https://github.com/libtcod/python-tcod/commit/60066f3}}
The developer fix directly passes the byte string \code{\_fmt(string)} as an argument to the function \code{lib.TCOD\_console\_printf\_ex}, avoiding the error caused by re-assigning the byte string to the variable \code{string}, which is previously annotated as type \code{str}.
The \name{}-suggested fix instead declares a new variable \code{byte\_string} for the byte string, and passes it to \code{lib.TCOD\_console\_printf\_ex} as an argument.

Figure~\ref{fig:cannotfix} shows a predicted fix that fails to remove the type error.\footnote{\href{https://github.com/networktocode/pyntc/commit/ebb35344e0121}{https://github.com/networktocode/pyntc/commit/ebb35344e0121}}
The developer fix uses a variable (\code{F5\_API\_DEVICE\_TYPE}) imported from another package. However, since the context code and the error message do not give any hint about the identifier to use, the model simply replaces it with \code{DEVICE\_TYPE}.
Because \name{} validates that a fix candidate removes the type error before reporting the fix to the user, this fix suggestion is not shown to users, highlighting the importance of validating fix candidates.

Finally, Figure~\ref{fig:unnaturalfix} fixes the type error but changes the semantics of the code in an unintended way.\footnote{\href{https://www.github.com/lykoss/lykos/commit/290f6e0d75e82eb8810106b5240b033c9abbd35c}{https://www.github.com/lykoss/lykos/commit/abbd35c}}
The error is because \code{Bot} and \code{new}, which is a variable, have incompatible types.
The developer fixes the error by asserting that \code{new} is of type \code{BotUser}.
\name{} instead suggests a fix that declares a new variable \code{new\_Bot}, which however fails to update the global \code{Bot} variable.
We include this example to show that \name{} is limited by relying on the type checker as the only validation mechanism.
Future work could address this limitation by further validating fixes by running a test suite.

\subsubsection{Type Fixes in the Wild}
To further validate the usefulness of \name{} in practice, we create pull requests with \name{}-suggested fixes for type errors. We run Pyre on different GitHub projects randomly picked among the projects in \dataset{}. 
In total, we create 30 pull requests (for 17 incompatible variable type errors, ten incompatible parameter type errors, and three invalid type errors).
By the time of this writing, \pr{} of the pull requests have been merged, six are still open, and four are closed.
For the pull requests merged so far, the developers generally were grateful about the changes.
In one case, the developers even asked us to apply similar fixes in other code locations, which we did, as we could use \name{}-suggested fixes there as well.
The four closed pull requests are: (i) two cases where the developers prefer to use type casts and dynamic type checks rather than updating the type annotations; (ii)~one case where the developers decided to suppress a warning about an incompatible \code{Optional} variable type; and (iii)~one case where the developers consider a warning about an incompatibility between \code{List[Optional[Path]]} and \code{List[None]} to be a false positive.
Overall, the developers' feedback confirms \name{}'s usefulness in practice.

\begin{finding}{RQ2}
\name{} successfully removes the type error in 85.4\% of the cases evaluated, and it finds exactly the developer-provided fix for 54.4\% of all errors.
\end{finding}

\subsection{RQ3: Ablation Study of \name{}}
\label{sec:evalablation}

\begin{table}[]
	\centering
	\caption{Ablation study and comparison with LLMs.}
	\setlength{\tabcolsep}{3pt}
	\label{table:resultablation}
		\begin{tabular}{@{}lrrr|rrr@{}}
			\toprule
			& \multicolumn{3}{c}{Error removal (\%)} & \multicolumn{3}{c}{Exact match (\%)} \\ \cmidrule(lr){2-4}\cmidrule(lr){5-7}
			Approach           & \multicolumn{1}{l}{Top-1} & Top-5         & Top-50         & \multicolumn{1}{l}{Top-1} & \multicolumn{1}{l}{Top-5} & \multicolumn{1}{l}{Top-50} \\ \midrule
			No pre-training        & 47.3        & 57.3        & 71.2  & 30.2       & 45.2       & 48.8     \\
			Vanilla TFix          & 4.6         & 11.0         & 16.7  & 0.0        & 1.1        & 1.8   \\ 
			No preprocessing         & 17.8        & 23.5        & 29.5   & 37.0       & 45.6       & 54.1    \\
			Small TFix model      & 43.1       & 63.3      & 79.0       & 32.7        & 44.8        & 53.0       \\
			\midrule
			text-davinci-003        & 21.7        & 27.8        & 34.6  & 14.6        & 18.1        & 20.9     \\
			gpt-3.5-turbo          & 21.9         & 23.8         & 26.0  & 10.3        & 12.1        & 14.5   \\ 
			gpt-4         & 34.1         & 36.7         & 39.1  & 18.9        & 22.1        & 26.4   \\ 
			\midrule
			Full \name{} & \textbf{50.9}            & \textbf{66.2} & \textbf{85.4} & \textbf{37.7}             & \textbf{48.0}             & \textbf{54.4}             \\
			\bottomrule
		\end{tabular}%
\end{table}

We perform an ablation study to evaluate the effectiveness of \name{} in different configurations. The upper part of Table~\ref{table:resultablation} summarizes the results discussed in the following.

\paragraph{No pre-training}
We train the T5 model directly on \dataset{}, i.e., without pre-training the model on the JavaScript APR tasks. The purpose of this experiment is to check if the knowledge of fixing JavaScript errors helps in fixing Python type errors.
We use the same experimental setup as discussed in Section~\ref{sec:implementation}, except that training continues beyond 30 epochs because the evaluation loss keeps decreasing. We train the model for 100 epochs and pick the model with the lowest validation loss, which is at the 32nd epoch.
The results show that pre-training the model on the JavaScript repair task contributes significantly to its effectiveness.
For example, the top-1 exact match rate drops from 37.7\% to 30.2\% without pre-training.

\paragraph{Vanilla TFix}
We try to predict the fix with the original TFix model, i.e., without fine-tuning TFix with \dataset{}.
The purpose of this experiment is to check whether gathering a dataset of type errors is really necessary.
This experiment uses the t5-large (770M parameters) model of TFix because removing fine-tuning also removes the resource constraints that motivated us to use the t5-base model (220M parameters).
For this experiment, we do not preprocess the Python source code as the tokenizer of the TFix model is trained without the special tokens.
As shown in Table~\ref{table:resultablation}, the effectiveness drops dramatically, e.g., to only 1.8\% exact matches within the top-50 suggestions.
The reasons are (i) that Python and JavaScript have different syntax, i.e., it is unlikely for the model to output syntactically correct Python code, and (ii) that the TFix model is not trained to fix type errors.

\paragraph{No preprocessing}
We try to generate a fix without the preprocessing that adds indentation and dedentation tokens (Section~\ref{sec:preprocess}).
We use the same experimental setup as discussed in Section~\ref{sec:implementation}, but we remove the special tokens from the input and output code.
We find preprocessing to be important, as otherwise the error removal rate drops significantly, e.g., from 50.9\% to 17.8\% in the top-1 prediction.
For exact match accuracy, the decrease in effectiveness is less strong, but the exact match might not be equal to the actual developer fix, as we ignore the newline tokens and the number of whitespaces for the comparison.

\paragraph{Small TFix model}
To study the impact of the model size, we try to predict the fix by basing \name{} on the small TFix model (with only 60M parameters).
We use the same experimental setup as discussed in Section~\ref{sec:implementation}.
As the evaluation loss of this model keeps decreasing beyond the 30th epoch, we train the model for 100 epochs, which converges at the 47th epoch.
The effectiveness of \name{} is negatively affected by using a smaller model, e.g., with 43.1\% instead of 50.9\% top-1 error removal rate.
At the same time, the negative impact of the small model can be partially compensated by considering more fix suggestions: For example, the top-50 exact match rate is reduced only slightly from 54.4\% to 53.0\%.
These results show that \name{} could also be effective in a resource-constrained setup, such as a developer laptop instead of a server.

\begin{finding}{RQ3}
	The full \name{} outperforms simpler variants of the approach, showing that each of \name{}'s components contributes to its effectiveness.
\end{finding}

\subsection{RQ4: Comparison with Prior Work}

\subsubsection{RQ4a: \name{} vs.\ Large Language Models}
\label{sec:rq4}
\label{sec:eval llms}

Fixing type errors relates to general-purpose APR~\cite{cacm2019-program-repair}.
The following compares \name{} with large language models (LLMs), which have been shown to yield state of the art results~\cite{xia2023automated,xia2023keep,jiang2023impact}.
\name{} and LLMs fundamentally differ in the sense that \name{} is designed and fine-tuned specifically for type error repair, whereas LLMs are trained in a task-independent manner, but typically on much more data.


\paragraph{Experimental Setup} 
We compare \name{} with three recent models offered by OpenAI: \emph{text-davinci-003}, \emph{gpt-3.5-turbo}, and \emph{gpt-4}.
Our prompt consists of five parts: a description of the task, the buggy code snippet, the type checker's error message, the line containing the error, and a description of the expected output format.

\paragraph{Results}
The lower part of Table~\ref{table:resultablation} shows the effectiveness of different models.
\name{} clearly outperforms all LLMs in terms of error removal and finding the exact developer fix.
The gpt-4 model, as the most recent and largest model, is the most effective LLM.
The text-davinci-003 model is slightly more effective than gpt-3.5-turbo, which may be because the latter is optimized for chat.
Manually analyzing the successful fixes, we notice that the LLMs mostly fix those errors that can be fixed with a single-token edit. Instead, \name{} can fix more complex type errors. 

\begin{finding}{RQ4a}
	\name{} is more effective than prompting general-purpose LLMs (54.4\% vs.\ 26.4\%).
\end{finding}







\subsubsection{RQ4b: \name{} vs.\ PyTER}
\label{sec:eval pyter}

Instead of targeting statically detected type errors, the recent PyTER~\cite{Oh2022} approach repairs bugs that manifest through a \code{TypeError} exception.
For a comparison, consider the two subproblems that both approaches address.
Subproblem~1 is \emph{detecting a type error}, done by the static type checker in our approach and by observing a runtime exception in PyTER.
Subproblem~2 is \emph{fixing a detected type error}, done by a neural model in our approach and by applying a set of repair templates in PyTER.
How \name{} and PyTER address subproblem~1 differs fundamentally.
While static type errors manifest without running the code, revealing a runtime type error require tests cases or a production run that triggers the error.
Moreover, a single type-related problem may manifest at different locations.
For example, a function that returns an incorrect value will manifest as a static type error at the return statement, but as a runtime type error at a code location that uses the value.
Because of these differences, performing a direct, end-to-end comparison is neither possible nor meaningful.
Instead, we quantify the overlap of the two approaches in terms of the errors they address and the fixes that they could potentially find, which answers four questions.

\paragraph{PyTER on \dataset{}}
\emph{1) How many of the errors in \dataset{} manifest via a runtime type error?}
We pick a random sample of 30 of all 281 fixes in our test set and inspect their commit messages.
The inspection shows that for 16/30 fixes, the problem was certainly found via static type checking, e.g., because the message mentions the type checker, and for 27/30 fixes, the problem was certainly not found via a \code{TypeError} thrown at runtime.
\emph{2) How many of the type errors in \dataset{} are in the scope of PyTER's fix templates?}
The repair templates cover three kinds of fixes:
adding an \code{instanceof} check, adding a type conversion, e.g., via a call to \code{int()}, and adding code to catch and handle a \code{TypeError} exception.
We check for each type error in our test set whether PyTER's repair templates can be instantiated into the fix, which shows that 15/281 type errors are in scope for PyTER, whereas the remaining 266 errors are not covered by any repair template.
Examples of fixes that are out-of-scope for PyTER are:
(i)~fixes that change a value, e.g., by modifying a string \code{"a b c"} into an array of strings \code{["a", "b", "c"]},
(ii)~fixes that change a type annotation, e.g., from \code{T} to \code{Optional[T]}, and
(iii)~fixes that add a call to \code{typing.cast()}.
In summary, PyTER address only a small fraction of the type errors in our dataset.

\paragraph{\name{} on PyTER's dataset}
\emph{3) How many of the errors in PyTER's dataset manifest via a static type error?}
The Pyre type checker that \name{} builds on checks (partially) type-annotated code only.
Among the 93 errors in PyTER's dataset, 16 are in a type-annotated function, and hence, checked at all, but the type checker does not find the errors fixed by PyTER.
\emph{4) How many of the type errors in PyTER's dataset are in the scope of \name{}'s neural model?} 
Our approach focuses on single-hunk fixes where the type error location is inside the hunk that needs to be changed.
While these assumptions commonly hold for static type errors (Section~\ref{sec:preliminary}), only 11/93 errors in PyTER's dataset match our assumptions.
In summary, \name{} addresses only a small fraction of the type errors in the PyTER dataset.

\begin{finding}{RQ4b}
	Our approach and PyTER~\cite{Oh2022} are complementary in the sense that they address type errors that manifest in different ways and that they apply different kinds of fixes.
\end{finding}

	\section{Discussion and Threats to Validity}

\paragraph{Python repositories} We select popular projects for our dataset, because recent work finds such projects to contain type annotations and type errors~\cite{Digrazia2022}.
A different set of repositories could yield different results, in particular for the preliminary study (Section~\ref{sec:preliminary}).

\paragraph{Limitations of static type checking}
\name{} builds upon the Pyre type checker, which, as all static type checkers, may suffer from false positives and false negatives.
A false positive, where the type checker incorrectly reports a type error in correct code, may lead to unnecessary code modifications by \name{}.
Conversely, a false negative, where an error goes unnoticed by the type checker, may cause \name{} to suggest a fix that does not really solve the problem, or even worse, introduces a new problem.
As a lower bound on \name{}'s effectiveness despite these limitations, we find that 54.4\% of the predicted fixes exactly match the developer's fix.
Other type checkers than Pyre may find different kinds of type errors and provide different kinds of hints for fixing them.
Because our approach uses the type checker as a black-box, adapting our implementation to support another type checker seems straightforward.

\paragraph{Type annotations}
Because the type checker reports errors only in functions that are at least partially type-annotated, \name{} cannot fix errors in completely unannotated code.
Despite this limitation, there is evidence that more and more code gets type-annotated, and hence, is in scope for \name{}.
For example, a recent study on the evolution of type annotations~\cite{Digrazia2022} finds 50 type annotations per 1,000 lines of code and an increasing trend on the adoption of type annotations.
Moreover, our dataset of thousands of real-world commits that address type errors shows that developers care about such errors.
Finally, as described in Section~1, large companies, such as Google, Dropbox, and Meta, are actively working toward type-annotating their Python code bases. 

\paragraph{Type errors}
\dataset{}, containing 2,766 real-world type error fixes, is filtered to contain only errors fixable with a single-hunk code change, and we cannot draw any conclusions about more complex fixes.
As shown in Section~\ref{sec:fix location}, many real-world fixes are local edits, which has motivated our design decision to focus on single-hunk fixes.
The distribution of error classes in \dataset{} reflects the errors that occur in practice, and does not cover all error classes that the type checker may find.
Thanks to the data-driven design of \name{}, the approach should be able to fix further classes of type errors when given corresponding training data.

\paragraph{Future work}
We plan to improve the error localization and will try different prompts to improve the performance of LLMs. Moreover, we plan to fine-tune different models beside TFix and apply \name{} to more classes of type errors. Finally, we plan to integrate our approach into an IDE.

\section{Related Work}

\paragraph{Automated program repair}
Earlier APR approaches~\cite{cacm2019-program-repair} can be classified into heuristic repair, e.g., based on the generate-and-validate method~\cite{Kim2013,LeGoues2012}, and constraint-based repair, which synthesizes a patch based on constraints~\cite{Nguyen2013b,Mechtaev2016}.
Both techniques rely on test suites, and hence, may suffer from overfitting~\cite{Qi2015}.
\name{} belongs to a more recent stream of work on learning-based repair.
In contrast to the above techniques, \name{} does not require tests but uses a static type checker to validate candidate fixes.
Other learning-based approaches include DrRepair~\cite{Yasunaga2020}, which fixes C compilation errors, Hoppity~\cite{dinella2020hoppity}, which represents fixes as a sequence of graph edits, and
Recoder~\cite{Zhu2021}, based on TreeGen~\cite{Sun2020TreeGen}, which proposes a syntax-guided edit decoder.
Compared to these GNN-based models, our approach uses a text-to-text transformer, which is easy to apply to any language.
Other text-based models include SequenceR~\cite{Chen2019} and work by Tufano et al.~\cite{Tufano2019}.
Vasic et al.~\cite{Vasic2019} propose to jointly localize and repair bugs. 
These approaches neither benefit from pre-training nor target Python type errors.
CoCoNut~\cite{Lutellier2020} combines multiple models using ensemble learning.
Instead, our approach learns how to fix all error types in one model.
Ye et al.\ incorporate feedback from compiling and executing tests to train a repair model~\cite{Ye2022a}, an idea that could also be adapted to type error repair.
Finally, motivated by recent results that show general-purpose LLMs to provide competitive results~\cite{DBLP:conf/sigsoft/XiaZ22,xia2023automated,xia2023keep,jiang2023impact}, we empirically compare \name{} with three LLMs (Section~\ref{sec:eval llms}).

We are aware of two APR approaches that target type errors.
Rite~\cite{Sakkas2020} is a template-based, data-driven approach for type errors in OCaml.
Their approach builds on a specifically designed, AST-based representation of fixes, while our approach uses textual inputs and outputs.
%
PyTER~\cite{Oh2022} is a test-based APR approach to fix runtime type errors in Python, which we empirically compare with in Section~\ref{sec:eval pyter}.
Their work and ours address related but ultimately different problems: PyTER requires test cases that trigger a runtime type error, but tests may not exist at all or have low coverage (e.g., \citet{Gruber2021} report a median coverage of 3.7\% across 22k Python projects).
In contrast, our work addresses statically detectable errors, and hence, is limited to errors that are statically detectable.
In practice, we expect PyTER and \name{} to complement each other.
Beyond type errors, several techniques for fixing other kinds of static analysis warnings have been proposed~\cite{Etemadi2023, oopsla2019, Marcillo2020}, which are also complementary to our work.

\paragraph{Type annotations and type errors} 
Several techniques predict types via deep learning~\cite{Hellendoorn2018,icse2019,Allamanis2020,Mir2021}, sometimes augmented with search-based validation of predicted types~\cite{fse2020} or static type inference~\cite{Peng2021}.
While these approaches focus on predicting correct types, \name{} addresses the complementary problem of fixing type-related errors.
\citet{Digrazia2022} show that adding type annotations to a code base often reveals statically detectable type errors, but that developers often do not find time to fix these errors.
Other studies investigate the impact of using type checkers in Python~\cite{Khan2021},
compare different type checkers with each other~\cite{Rak-amnouykit2020}, and 
provide evidence that static typing may reduce the bug fixing effort~\cite{Zhang2021bugs}.
Both \citet{Digrazia2022} and \citet{Rak-amnouykit2020} report that type-annotated repositories rarely type-check, showing the need for an APR tool for Python type errors, such as \name{}.

\paragraph{Transfer learning on code} Pre-trained models for source code, e.g., CodeBERT~\cite{Feng2020}, GraphCodeBERT~\cite{Guo2020b}, CodeT5~\cite{Wang2021} and CodeTrans~\cite{Elnaggar2021}, achieve promising results~\cite{Wan2022}.
Following this paradigm, we build on TFix because it is already trained on an APR task. 
Another work that applies transfer learning in language models of code is VRepair~\cite{Chen2022Vrepair}. They pre-train a transformer model on a large bug fix dataset for C, and then fine-tune it with a vulnerability fix dataset for C.
Our work shows that the benefits of transferring knowledge not only between different fixing tasks, but also between different programming languages.

\paragraph{Delta debugging}
Delta debugging~\cite{zeller2002isolating} finds failure-inducing code hunks in a commit, and is widely used, e.g., for fault localization~\cite{wong2016survey}. We treat hunks that fix type errors as ``failure-inducing'', which is similar to prior work~\cite{bavishi2019phoenix} but adopted to type errors.

	\section{Conclusion}

This paper presents \name{}, the first automated repair technique targeted specifically at statically detectable type errors in Python.
The design of the approach is motivated by the findings of a preliminary study.
To generate a relevant dataset, we apply a combination of delta debugging and type checking, which results in \dataset{}, containing 2,766 Python type errors and fixes.
We then present cross-lingual transfer learning, which addresses the problem of having a small dataset for a deep learning model by fine-tuning an existing APR model originally trained for another task and language. 
Our evaluation shows the effectiveness of \name{}, e.g., by providing a fix that removes the targeted type error for 85.4\% of the studied errors. Finally, as of this writing, \pr{} out of 30 GitHub pull requests based on \name{}-generated type error fixes have been merged by developers, demonstrating the usefulness of \name{} in practice.


\section*{Data Availability}

\noindent
The \dataset{} dataset, the \name{} implementation, and our experimental results
are available online: \url{https://github.com/sola-st/PyTy} and \url{https://zenodo.org/records/10441045}.

\section*{Acknowledgment}
This work was supported by the European Research Council (ERC, grant agreement 851895), and by the German Research Foundation within the ConcSys and DeMoCo projects. We would like to
thank the reviewers for their valuable feedback to improve
the paper.
	
	\newpage

	\bibliographystyle{ACM-Reference-Format}
	\bibliography{references,referencesMP}


\begin{thebibliography}{56}


\ifx \showCODEN    \undefined \def \showCODEN     #1{\unskip}     \fi
\ifx \showDOI      \undefined \def \showDOI       #1{#1}\fi
\ifx \showISBNx    \undefined \def \showISBNx     #1{\unskip}     \fi
\ifx \showISBNxiii \undefined \def \showISBNxiii  #1{\unskip}     \fi
\ifx \showISSN     \undefined \def \showISSN      #1{\unskip}     \fi
\ifx \showLCCN     \undefined \def \showLCCN      #1{\unskip}     \fi
\ifx \shownote     \undefined \def \shownote      #1{#1}          \fi
\ifx \showarticletitle \undefined \def \showarticletitle #1{#1}   \fi
\ifx \showURL      \undefined \def \showURL       {\relax}        \fi
\providecommand\bibfield[2]{#2}
\providecommand\bibinfo[2]{#2}
\providecommand\natexlab[1]{#1}
\providecommand\showeprint[2][]{arXiv:#2}

\bibitem[Allamanis et~al\mbox{.}(2020)]%
        {Allamanis2020}
\bibfield{author}{\bibinfo{person}{Miltiadis Allamanis},
  \bibinfo{person}{Earl~T. Barr}, \bibinfo{person}{Soline Ducousso}, {and}
  \bibinfo{person}{Zheng Gao}.} \bibinfo{year}{2020}\natexlab{}.
\newblock \showarticletitle{Typilus: Neural Type Hints}. In
  \bibinfo{booktitle}{\emph{PLDI}}.
\newblock


\bibitem[Allamanis et~al\mbox{.}(2018)]%
        {allamanis2017learning}
\bibfield{author}{\bibinfo{person}{Miltiadis Allamanis}, \bibinfo{person}{Marc
  Brockschmidt}, {and} \bibinfo{person}{Mahmoud Khademi}.}
  \bibinfo{year}{2018}\natexlab{}.
\newblock \showarticletitle{Learning to Represent Programs with Graphs}. In
  \bibinfo{booktitle}{\emph{International Conference on Learning
  Representations (ICLR)}}.
\newblock


\bibitem[Bader et~al\mbox{.}(2019)]%
        {oopsla2019}
\bibfield{author}{\bibinfo{person}{Johannes Bader}, \bibinfo{person}{Andrew
  Scott}, \bibinfo{person}{Michael Pradel}, {and} \bibinfo{person}{Satish
  Chandra}.} \bibinfo{year}{2019}\natexlab{}.
\newblock \showarticletitle{Getafix: Learning to Fix Bugs Automatically}. In
  \bibinfo{booktitle}{\emph{OOPSLA}}. \bibinfo{pages}{159:1--159:27}.
\newblock


\bibitem[Bavishi et~al\mbox{.}(2019)]%
        {bavishi2019phoenix}
\bibfield{author}{\bibinfo{person}{Rohan Bavishi}, \bibinfo{person}{Hiroaki
  Yoshida}, {and} \bibinfo{person}{Mukul~R Prasad}.}
  \bibinfo{year}{2019}\natexlab{}.
\newblock \showarticletitle{Phoenix: Automated data-driven synthesis of repairs
  for static analysis violations}. In \bibinfo{booktitle}{\emph{Proceedings of
  the 2019 27th ACM Joint Meeting on European Software Engineering Conference
  and Symposium on the Foundations of Software Engineering}}.
  \bibinfo{pages}{613--624}.
\newblock


\bibitem[Berabi et~al\mbox{.}(2021)]%
        {Berabi2021}
\bibfield{author}{\bibinfo{person}{Berkay Berabi}, \bibinfo{person}{Jingxuan
  He}, \bibinfo{person}{Veselin Raychev}, {and} \bibinfo{person}{Martin
  Vechev}.} \bibinfo{year}{2021}\natexlab{}.
\newblock \showarticletitle{TFix: Learning to Fix Coding Errors with a
  Text-to-Text Transformer}. In \bibinfo{booktitle}{\emph{ICML}}.
\newblock


\bibitem[Chen et~al\mbox{.}(2019)]%
        {Chen2019}
\bibfield{author}{\bibinfo{person}{Zimin Chen}, \bibinfo{person}{Steve
  Kommrusch}, \bibinfo{person}{Michele Tufano},
  \bibinfo{person}{Louis{-}No{\"{e}}l Pouchet}, \bibinfo{person}{Denys
  Poshyvanyk}, {and} \bibinfo{person}{Martin Monperrus}.}
  \bibinfo{year}{2019}\natexlab{}.
\newblock \showarticletitle{SequenceR: Sequence-to-Sequence Learning for
  End-to-End Program Repair}.
\newblock \bibinfo{journal}{\emph{IEEE TSE}} (\bibinfo{year}{2019}).
\newblock


\bibitem[Chen et~al\mbox{.}(2022)]%
        {Chen2022Vrepair}
\bibfield{author}{\bibinfo{person}{Zimin Chen}, \bibinfo{person}{Steve~James
  Kommrusch}, {and} \bibinfo{person}{Martin Monperrus}.}
  \bibinfo{year}{2022}\natexlab{}.
\newblock \showarticletitle{Neural Transfer Learning for Repairing Security
  Vulnerabilities in C Code}.
\newblock \bibinfo{journal}{\emph{IEEE Transactions on Software Engineering}}
  (\bibinfo{year}{2022}), \bibinfo{pages}{1--1}.
\newblock
\urldef\tempurl%
\url{https://doi.org/10.1109/TSE.2022.3147265}
\showDOI{\tempurl}


\bibitem[Cohen(1960)]%
        {cohen1960coefficient}
\bibfield{author}{\bibinfo{person}{Jacob Cohen}.}
  \bibinfo{year}{1960}\natexlab{}.
\newblock \showarticletitle{A coefficient of agreement for nominal scales}.
\newblock \bibinfo{journal}{\emph{Educational and psychological measurement}}
  \bibinfo{volume}{20}, \bibinfo{number}{1} (\bibinfo{year}{1960}),
  \bibinfo{pages}{37--46}.
\newblock


\bibitem[{Di Grazia} and Pradel(2022)]%
        {Digrazia2022}
\bibfield{author}{\bibinfo{person}{Luca {Di Grazia}} {and}
  \bibinfo{person}{Michael Pradel}.} \bibinfo{year}{2022}\natexlab{}.
\newblock \showarticletitle{The Evolution of Type Annotations in Python: An
  Empirical Study}. In \bibinfo{booktitle}{\emph{Proceedings of the 30th ACM
  Joint European Software Engineering Conference and Symposium on the
  Foundations of Software Engineering}} (Singapore, Singapore)
  \emph{(\bibinfo{series}{ESEC/FSE 2022})}. \bibinfo{publisher}{Association for
  Computing Machinery}, \bibinfo{address}{New York, NY, USA},
  \bibinfo{pages}{209–220}.
\newblock
\showISBNx{9781450394130}
\urldef\tempurl%
\url{https://doi.org/10.1145/3540250.3549114}
\showDOI{\tempurl}


\bibitem[Dinella et~al\mbox{.}(2020)]%
        {dinella2020hoppity}
\bibfield{author}{\bibinfo{person}{Elizabeth Dinella}, \bibinfo{person}{Hanjun
  Dai}, \bibinfo{person}{Ziyang Li}, \bibinfo{person}{Mayur Naik},
  \bibinfo{person}{Le Song}, {and} \bibinfo{person}{Ke Wang}.}
  \bibinfo{year}{2020}\natexlab{}.
\newblock \showarticletitle{Hoppity: Learning graph transformations to detect
  and fix bugs in programs}. In \bibinfo{booktitle}{\emph{International
  Conference on Learning Representations (ICLR)}}.
\newblock


\bibitem[Elnaggar et~al\mbox{.}(2021)]%
        {Elnaggar2021}
\bibfield{author}{\bibinfo{person}{Ahmed Elnaggar}, \bibinfo{person}{Wei Ding},
  \bibinfo{person}{Llion Jones}, \bibinfo{person}{Tom Gibbs},
  \bibinfo{person}{Tamas Feher}, \bibinfo{person}{Christoph Angerer},
  \bibinfo{person}{Silvia Severini}, \bibinfo{person}{Florian Matthes}, {and}
  \bibinfo{person}{Burkhard Rost}.} \bibinfo{year}{2021}\natexlab{}.
\newblock \showarticletitle{CodeTrans: Towards Cracking the Language of
  Silicon's Code Through Self-Supervised Deep Learning and High Performance
  Computing}.
\newblock \bibinfo{journal}{\emph{arXiv preprint arXiv:2104.02443}}
  (\bibinfo{year}{2021}).
\newblock


\bibitem[Etemadi et~al\mbox{.}(2023)]%
        {Etemadi2023}
\bibfield{author}{\bibinfo{person}{Khashayar Etemadi}, \bibinfo{person}{Nicolas
  Harrand}, \bibinfo{person}{Simon Larsén}, \bibinfo{person}{Haris Adzemovic},
  \bibinfo{person}{Henry~Luong Phu}, \bibinfo{person}{Ashutosh Verma},
  \bibinfo{person}{Fernanda Madeiral}, \bibinfo{person}{Douglas Wikström},
  {and} \bibinfo{person}{Martin Monperrus}.} \bibinfo{year}{2023}\natexlab{}.
\newblock \showarticletitle{Sorald: Automatic Patch Suggestions for SonarQube
  Static Analysis Violations}.
\newblock \bibinfo{journal}{\emph{IEEE Transactions on Dependable and Secure
  Computing}} \bibinfo{volume}{20}, \bibinfo{number}{4} (\bibinfo{year}{2023}),
  \bibinfo{pages}{2794--2810}.
\newblock
\urldef\tempurl%
\url{https://doi.org/10.1109/TDSC.2022.3167316}
\showDOI{\tempurl}


\bibitem[Feng et~al\mbox{.}(2020)]%
        {Feng2020}
\bibfield{author}{\bibinfo{person}{Zhangyin Feng}, \bibinfo{person}{Daya Guo},
  \bibinfo{person}{Duyu Tang}, \bibinfo{person}{Nan Duan},
  \bibinfo{person}{Xiaocheng Feng}, \bibinfo{person}{Ming Gong},
  \bibinfo{person}{Linjun Shou}, \bibinfo{person}{Bing Qin},
  \bibinfo{person}{Ting Liu}, \bibinfo{person}{Daxin Jiang}, {and}
  \bibinfo{person}{Ming Zhou}.} \bibinfo{year}{2020}\natexlab{}.
\newblock \showarticletitle{CodeBERT: {A} Pre-Trained Model for Programming and
  Natural Languages}. In \bibinfo{booktitle}{\emph{Findings of the Association
  for Computational Linguistics: {EMNLP} 2020, Online Event, 16-20 November
  2020}} \emph{(\bibinfo{series}{Findings of {ACL}},
  Vol.~\bibinfo{volume}{{EMNLP} 2020})},
  \bibfield{editor}{\bibinfo{person}{Trevor Cohn}, \bibinfo{person}{Yulan He},
  {and} \bibinfo{person}{Yang Liu}} (Eds.). \bibinfo{publisher}{Association for
  Computational Linguistics}, \bibinfo{pages}{1536--1547}.
\newblock
\urldef\tempurl%
\url{https://doi.org/10.18653/v1/2020.findings-emnlp.139}
\showDOI{\tempurl}


\bibitem[Glaser et~al\mbox{.}(1968)]%
        {glaser1968discovery}
\bibfield{author}{\bibinfo{person}{Barney~G Glaser}, \bibinfo{person}{Anselm~L
  Strauss}, {and} \bibinfo{person}{Elizabeth Strutzel}.}
  \bibinfo{year}{1968}\natexlab{}.
\newblock \showarticletitle{The discovery of grounded theory; strategies for
  qualitative research}.
\newblock \bibinfo{journal}{\emph{Nursing research}} \bibinfo{volume}{17},
  \bibinfo{number}{4} (\bibinfo{year}{1968}), \bibinfo{pages}{364}.
\newblock


\bibitem[Gruber et~al\mbox{.}(2021)]%
        {Gruber2021}
\bibfield{author}{\bibinfo{person}{Martin Gruber}, \bibinfo{person}{Stephan
  Lukasczyk}, \bibinfo{person}{Florian Kroi{\ss}}, {and}
  \bibinfo{person}{Gordon Fraser}.} \bibinfo{year}{2021}\natexlab{}.
\newblock \showarticletitle{An empirical study of flaky tests in python}. In
  \bibinfo{booktitle}{\emph{2021 14th IEEE Conference on Software Testing,
  Verification and Validation (ICST)}}. IEEE, \bibinfo{pages}{148--158}.
\newblock


\bibitem[Guo et~al\mbox{.}(2021)]%
        {Guo2020b}
\bibfield{author}{\bibinfo{person}{Daya Guo}, \bibinfo{person}{Shuo Ren},
  \bibinfo{person}{Shuai Lu}, \bibinfo{person}{Zhangyin Feng},
  \bibinfo{person}{Duyu Tang}, \bibinfo{person}{Shujie Liu},
  \bibinfo{person}{Long Zhou}, \bibinfo{person}{Nan Duan},
  \bibinfo{person}{Alexey Svyatkovskiy}, \bibinfo{person}{Shengyu Fu},
  \bibinfo{person}{Michele Tufano}, \bibinfo{person}{Shao~Kun Deng},
  \bibinfo{person}{Colin~B. Clement}, \bibinfo{person}{Dawn Drain},
  \bibinfo{person}{Neel Sundaresan}, \bibinfo{person}{Jian Yin},
  \bibinfo{person}{Daxin Jiang}, {and} \bibinfo{person}{Ming Zhou}.}
  \bibinfo{year}{2021}\natexlab{}.
\newblock \showarticletitle{GraphCodeBERT: Pre-training Code Representations
  with Data Flow}. In \bibinfo{booktitle}{\emph{9th International Conference on
  Learning Representations, {ICLR} 2021, Virtual Event, Austria, May 3-7,
  2021}}. \bibinfo{publisher}{OpenReview.net}.
\newblock
\urldef\tempurl%
\url{https://openreview.net/forum?id=jLoC4ez43PZ}
\showURL{%
\tempurl}


\bibitem[Hellendoorn et~al\mbox{.}(2018)]%
        {Hellendoorn2018}
\bibfield{author}{\bibinfo{person}{Vincent~J. Hellendoorn},
  \bibinfo{person}{Christian Bird}, \bibinfo{person}{Earl~T. Barr}, {and}
  \bibinfo{person}{Miltiadis Allamanis}.} \bibinfo{year}{2018}\natexlab{}.
\newblock \showarticletitle{Deep learning type inference}. In
  \bibinfo{booktitle}{\emph{Proceedings of the 2018 {ACM} Joint Meeting on
  European Software Engineering Conference and Symposium on the Foundations of
  Software Engineering, {ESEC/SIGSOFT} {FSE} 2018, Lake Buena Vista, FL, USA,
  November 04-09, 2018}}, \bibfield{editor}{\bibinfo{person}{Gary~T. Leavens},
  \bibinfo{person}{Alessandro Garcia}, {and} \bibinfo{person}{Corina~S.
  Pasareanu}} (Eds.). \bibinfo{publisher}{{ACM}}, \bibinfo{pages}{152--162}.
\newblock
\urldef\tempurl%
\url{https://doi.org/10.1145/3236024.3236051}
\showDOI{\tempurl}


\bibitem[Herfert et~al\mbox{.}(2017)]%
        {ase2017-GTR}
\bibfield{author}{\bibinfo{person}{Satia Herfert}, \bibinfo{person}{Jibesh
  Patra}, {and} \bibinfo{person}{Michael Pradel}.}
  \bibinfo{year}{2017}\natexlab{}.
\newblock \showarticletitle{Automatically Reducing Tree-Structured Test
  Inputs}. In \bibinfo{booktitle}{\emph{ASE}}.
\newblock


\bibitem[Jiang et~al\mbox{.}(2023)]%
        {jiang2023impact}
\bibfield{author}{\bibinfo{person}{Nan Jiang}, \bibinfo{person}{Kevin Liu},
  \bibinfo{person}{Thibaud Lutellier}, {and} \bibinfo{person}{Lin Tan}.}
  \bibinfo{year}{2023}\natexlab{}.
\newblock \showarticletitle{Impact of code language models on automated program
  repair}. In \bibinfo{booktitle}{\emph{Proceedings of the 45th International
  Conference on Software Engineering (ICSE 2023). Association for Computing
  Machinery}}.
\newblock


\bibitem[Khan et~al\mbox{.}(2021)]%
        {Khan2021}
\bibfield{author}{\bibinfo{person}{Faizan Khan}, \bibinfo{person}{Boqi Chen},
  \bibinfo{person}{Daniel Varro}, {and} \bibinfo{person}{Shane Mcintosh}.}
  \bibinfo{year}{2021}\natexlab{}.
\newblock \showarticletitle{An Empirical Study of Type-Related Defects in
  Python Projects}.
\newblock \bibinfo{journal}{\emph{IEEE Transactions on Software Engineering}}
  (\bibinfo{year}{2021}).
\newblock


\bibitem[Kim et~al\mbox{.}(2013)]%
        {Kim2013}
\bibfield{author}{\bibinfo{person}{Dongsun Kim}, \bibinfo{person}{Jaechang
  Nam}, \bibinfo{person}{Jaewoo Song}, {and} \bibinfo{person}{Sunghun Kim}.}
  \bibinfo{year}{2013}\natexlab{}.
\newblock \showarticletitle{Automatic patch generation learned from
  human-written patches.}. In \bibinfo{booktitle}{\emph{International
  Conference on Software Engineering (ICSE)}}. \bibinfo{pages}{802--811}.
\newblock


\bibitem[Kudo and Richardson(2018)]%
        {kudo2018sentencepiece}
\bibfield{author}{\bibinfo{person}{Taku Kudo} {and} \bibinfo{person}{John
  Richardson}.} \bibinfo{year}{2018}\natexlab{}.
\newblock \showarticletitle{SentencePiece: A simple and language independent
  subword tokenizer and detokenizer for Neural Text Processing}. In
  \bibinfo{booktitle}{\emph{Conference on Empirical Methods in Natural Language
  Processing}}.
\newblock


\bibitem[Landis and Koch(1977)]%
        {Landis1977}
\bibfield{author}{\bibinfo{person}{J.~Richard Landis} {and}
  \bibinfo{person}{Gary~G. Koch}.} \bibinfo{year}{1977}\natexlab{}.
\newblock \showarticletitle{The Measurement of Observer Agreement for
  Categorical Data}.
\newblock \bibinfo{journal}{\emph{Biometrics}} \bibinfo{volume}{33},
  \bibinfo{number}{1} (\bibinfo{year}{1977}), \bibinfo{pages}{159--174}.
\newblock
\showISSN{0006341X, 15410420}
\urldef\tempurl%
\url{http://www.jstor.org/stable/2529310}
\showURL{%
\tempurl}


\bibitem[{Le Goues} et~al\mbox{.}(2012)]%
        {LeGoues2012}
\bibfield{author}{\bibinfo{person}{Claire {Le Goues}}, \bibinfo{person}{ThanhVu
  Nguyen}, \bibinfo{person}{Stephanie Forrest}, {and} \bibinfo{person}{Westley
  Weimer}.} \bibinfo{year}{2012}\natexlab{}.
\newblock \showarticletitle{GenProg: {A} Generic Method for Automatic Software
  Repair}.
\newblock \bibinfo{journal}{\emph{{IEEE} Trans. Software Eng.}}
  \bibinfo{volume}{38}, \bibinfo{number}{1} (\bibinfo{year}{2012}),
  \bibinfo{pages}{54--72}.
\newblock


\bibitem[{Le Goues} et~al\mbox{.}(2019)]%
        {cacm2019-program-repair}
\bibfield{author}{\bibinfo{person}{Claire {Le Goues}}, \bibinfo{person}{Michael
  Pradel}, {and} \bibinfo{person}{Abhik Roychoudhury}.}
  \bibinfo{year}{2019}\natexlab{}.
\newblock \showarticletitle{Automated program repair}.
\newblock \bibinfo{journal}{\emph{Commun. {ACM}}} \bibinfo{volume}{62},
  \bibinfo{number}{12} (\bibinfo{year}{2019}), \bibinfo{pages}{56--65}.
\newblock
\urldef\tempurl%
\url{https://doi.org/10.1145/3318162}
\showDOI{\tempurl}


\bibitem[Lutellier et~al\mbox{.}(2020)]%
        {Lutellier2020}
\bibfield{author}{\bibinfo{person}{Thibaud Lutellier},
  \bibinfo{person}{Hung~Viet Pham}, \bibinfo{person}{Lawrence Pang},
  \bibinfo{person}{Yitong Li}, \bibinfo{person}{Moshi Wei}, {and}
  \bibinfo{person}{Lin Tan}.} \bibinfo{year}{2020}\natexlab{}.
\newblock \showarticletitle{CoCoNuT: combining context-aware neural translation
  models using ensemble for program repair}. In
  \bibinfo{booktitle}{\emph{{ISSTA} '20: 29th {ACM} {SIGSOFT} International
  Symposium on Software Testing and Analysis, Virtual Event, USA, July 18-22,
  2020}}, \bibfield{editor}{\bibinfo{person}{Sarfraz Khurshid} {and}
  \bibinfo{person}{Corina~S. Pasareanu}} (Eds.). \bibinfo{publisher}{{ACM}},
  \bibinfo{pages}{101--114}.
\newblock
\urldef\tempurl%
\url{https://doi.org/10.1145/3395363.3397369}
\showDOI{\tempurl}


\bibitem[Malik et~al\mbox{.}(2019)]%
        {icse2019}
\bibfield{author}{\bibinfo{person}{Rabee~Sohail Malik}, \bibinfo{person}{Jibesh
  Patra}, {and} \bibinfo{person}{Michael Pradel}.}
  \bibinfo{year}{2019}\natexlab{}.
\newblock \showarticletitle{{NL2Type}: {I}nferring {JavaScript} function types
  from natural language information}. In \bibinfo{booktitle}{\emph{Proceedings
  of the 41st International Conference on Software Engineering, {ICSE} 2019,
  Montreal, QC, Canada, May 25-31, 2019}}. \bibinfo{pages}{304--315}.
\newblock
\urldef\tempurl%
\url{https://doi.org/10.1109/ICSE.2019.00045}
\showDOI{\tempurl}


\bibitem[Marcilio et~al\mbox{.}(2020)]%
        {Marcillo2020}
\bibfield{author}{\bibinfo{person}{Diego Marcilio}, \bibinfo{person}{Carlo~A.
  Furia}, \bibinfo{person}{Rodrigo Bonif{\'{a}}cio}, {and}
  \bibinfo{person}{Gustavo Pinto}.} \bibinfo{year}{2020}\natexlab{}.
\newblock \showarticletitle{SpongeBugs: Automatically generating fix
  suggestions in response to static code analysis warnings}.
\newblock \bibinfo{journal}{\emph{J. Syst. Softw.}}  \bibinfo{volume}{168}
  (\bibinfo{year}{2020}), \bibinfo{pages}{110671}.
\newblock
\urldef\tempurl%
\url{https://doi.org/10.1016/j.jss.2020.110671}
\showDOI{\tempurl}


\bibitem[Mechtaev et~al\mbox{.}(2016)]%
        {Mechtaev2016}
\bibfield{author}{\bibinfo{person}{Sergey Mechtaev}, \bibinfo{person}{Jooyong
  Yi}, {and} \bibinfo{person}{Abhik Roychoudhury}.}
  \bibinfo{year}{2016}\natexlab{}.
\newblock \showarticletitle{Angelix: Scalable multiline program patch synthesis
  via symbolic analysis}. In \bibinfo{booktitle}{\emph{Proceedings of the 38th
  international conference on software engineering}}.
  \bibinfo{pages}{691--701}.
\newblock


\bibitem[Mir et~al\mbox{.}(2022)]%
        {Mir2021}
\bibfield{author}{\bibinfo{person}{Amir~M. Mir}, \bibinfo{person}{Evaldas
  Lato\v{s}kinas}, \bibinfo{person}{Sebastian Proksch}, {and}
  \bibinfo{person}{Georgios Gousios}.} \bibinfo{year}{2022}\natexlab{}.
\newblock \showarticletitle{Type4Py: Practical Deep Similarity Learning-Based
  Type Inference for Python}. In \bibinfo{booktitle}{\emph{Proceedings of the
  44th International Conference on Software Engineering}} (Pittsburgh,
  Pennsylvania) \emph{(\bibinfo{series}{ICSE '22})}.
  \bibinfo{publisher}{Association for Computing Machinery},
  \bibinfo{address}{New York, NY, USA}, \bibinfo{pages}{2241–2252}.
\newblock
\showISBNx{9781450392211}
\urldef\tempurl%
\url{https://doi.org/10.1145/3510003.3510124}
\showDOI{\tempurl}


\bibitem[Nguyen et~al\mbox{.}(2013)]%
        {Nguyen2013b}
\bibfield{author}{\bibinfo{person}{Hoang Duong~Thien Nguyen},
  \bibinfo{person}{Dawei Qi}, \bibinfo{person}{Abhik Roychoudhury}, {and}
  \bibinfo{person}{Satish Chandra}.} \bibinfo{year}{2013}\natexlab{}.
\newblock \showarticletitle{SemFix: program repair via semantic analysis}. In
  \bibinfo{booktitle}{\emph{35th International Conference on Software
  Engineering, {ICSE} '13, San Francisco, CA, USA, May 18-26, 2013}}.
  \bibinfo{pages}{772--781}.
\newblock


\bibitem[Oh and Oh(2022)]%
        {Oh2022}
\bibfield{author}{\bibinfo{person}{Wonseok Oh} {and} \bibinfo{person}{Hakjoo
  Oh}.} \bibinfo{year}{2022}\natexlab{}.
\newblock \showarticletitle{PyTER: Effective Program Repair for Python Type
  Errors}. In \bibinfo{booktitle}{\emph{ESEC/FSE}}.
\newblock


\bibitem[Peng et~al\mbox{.}(2022)]%
        {Peng2021}
\bibfield{author}{\bibinfo{person}{Yun Peng}, \bibinfo{person}{Cuiyun Gao},
  \bibinfo{person}{Zongjie Li}, \bibinfo{person}{Bowei Gao},
  \bibinfo{person}{David Lo}, \bibinfo{person}{Qirun Zhang}, {and}
  \bibinfo{person}{Michael Lyu}.} \bibinfo{year}{2022}\natexlab{}.
\newblock \showarticletitle{Static Inference Meets Deep Learning: A Hybrid Type
  Inference Approach for Python}. In \bibinfo{booktitle}{\emph{Proceedings of
  the 44th International Conference on Software Engineering}} (Pittsburgh,
  Pennsylvania) \emph{(\bibinfo{series}{ICSE '22})}.
  \bibinfo{publisher}{Association for Computing Machinery},
  \bibinfo{address}{New York, NY, USA}, \bibinfo{pages}{2019–2030}.
\newblock
\showISBNx{9781450392211}
\urldef\tempurl%
\url{https://doi.org/10.1145/3510003.3510038}
\showDOI{\tempurl}


\bibitem[Pradel et~al\mbox{.}(2020)]%
        {fse2020}
\bibfield{author}{\bibinfo{person}{Michael Pradel}, \bibinfo{person}{Georgios
  Gousios}, \bibinfo{person}{Jason Liu}, {and} \bibinfo{person}{Satish
  Chandra}.} \bibinfo{year}{2020}\natexlab{}.
\newblock \showarticletitle{TypeWriter: Neural Type Prediction with
  Search-based Validation}. In \bibinfo{booktitle}{\emph{{ESEC/FSE} '20: 28th
  {ACM} Joint European Software Engineering Conference and Symposium on the
  Foundations of Software Engineering, Virtual Event, USA, November 8-13,
  2020}}. \bibinfo{pages}{209--220}.
\newblock
\urldef\tempurl%
\url{https://doi.org/10.1145/3368089.3409715}
\showURL{%
\tempurl}


\bibitem[Pradel and Sen(2018)]%
        {oopsla2018-DeepBugs}
\bibfield{author}{\bibinfo{person}{Michael Pradel} {and}
  \bibinfo{person}{Koushik Sen}.} \bibinfo{year}{2018}\natexlab{}.
\newblock \showarticletitle{{DeepBugs}: A learning approach to name-based bug
  detection}.
\newblock \bibinfo{journal}{\emph{{PACMPL}}} \bibinfo{volume}{2},
  \bibinfo{number}{{OOPSLA}} (\bibinfo{year}{2018}),
  \bibinfo{pages}{147:1--147:25}.
\newblock
\urldef\tempurl%
\url{https://doi.org/10.1145/3276517}
\showURL{%
\tempurl}


\bibitem[Qi et~al\mbox{.}(2015)]%
        {Qi2015}
\bibfield{author}{\bibinfo{person}{Zichao Qi}, \bibinfo{person}{Fan Long},
  \bibinfo{person}{Sara Achour}, {and} \bibinfo{person}{Martin Rinard}.}
  \bibinfo{year}{2015}\natexlab{}.
\newblock \showarticletitle{An analysis of patch plausibility and correctness
  for generate-and-validate patch generation systems}. In
  \bibinfo{booktitle}{\emph{Proceedings of the 2015 International Symposium on
  Software Testing and Analysis}}. \bibinfo{pages}{24--36}.
\newblock


\bibitem[Raffel et~al\mbox{.}(2020)]%
        {Raffel2020}
\bibfield{author}{\bibinfo{person}{Colin Raffel}, \bibinfo{person}{Noam
  Shazeer}, \bibinfo{person}{Adam Roberts}, \bibinfo{person}{Katherine Lee},
  \bibinfo{person}{Sharan Narang}, \bibinfo{person}{Michael Matena},
  \bibinfo{person}{Yanqi Zhou}, \bibinfo{person}{Wei Li}, {and}
  \bibinfo{person}{Peter~J. Liu}.} \bibinfo{year}{2020}\natexlab{}.
\newblock \showarticletitle{Exploring the Limits of Transfer Learning with a
  Unified Text-to-Text Transformer}.
\newblock \bibinfo{journal}{\emph{J. Mach. Learn. Res.}}  \bibinfo{volume}{21}
  (\bibinfo{year}{2020}), \bibinfo{pages}{140:1--140:67}.
\newblock
\urldef\tempurl%
\url{http://jmlr.org/papers/v21/20-074.html}
\showURL{%
\tempurl}


\bibitem[Rak-amnouykit et~al\mbox{.}(2020)]%
        {Rak-amnouykit2020}
\bibfield{author}{\bibinfo{person}{Ingkarat Rak-amnouykit},
  \bibinfo{person}{Daniel McCrevan}, \bibinfo{person}{Ana Milanova},
  \bibinfo{person}{Martin Hirzel}, {and} \bibinfo{person}{Julian Dolby}.}
  \bibinfo{year}{2020}\natexlab{}.
\newblock \showarticletitle{Python 3 Types in the Wild: A Tale of Two Type
  Systems}. In \bibinfo{booktitle}{\emph{DLS}}.
\newblock


\bibitem[Sakkas et~al\mbox{.}(2020)]%
        {Sakkas2020}
\bibfield{author}{\bibinfo{person}{Georgios Sakkas}, \bibinfo{person}{Madeline
  Endres}, \bibinfo{person}{Benjamin Cosman}, \bibinfo{person}{Westley Weimer},
  {and} \bibinfo{person}{Ranjit Jhala}.} \bibinfo{year}{2020}\natexlab{}.
\newblock \showarticletitle{Type error feedback via analytic program repair}.
  In \bibinfo{booktitle}{\emph{Proceedings of the 41st {ACM} {SIGPLAN}
  International Conference on Programming Language Design and Implementation,
  {PLDI} 2020, London, UK, June 15-20, 2020}},
  \bibfield{editor}{\bibinfo{person}{Alastair~F. Donaldson} {and}
  \bibinfo{person}{Emina Torlak}} (Eds.). \bibinfo{publisher}{{ACM}},
  \bibinfo{pages}{16--30}.
\newblock
\urldef\tempurl%
\url{https://doi.org/10.1145/3385412.3386005}
\showDOI{\tempurl}


\bibitem[Siek and Taha(2007)]%
        {DBLP:conf/ecoop/SiekT07}
\bibfield{author}{\bibinfo{person}{Jeremy~G. Siek} {and} \bibinfo{person}{Walid
  Taha}.} \bibinfo{year}{2007}\natexlab{}.
\newblock \showarticletitle{Gradual Typing for Objects}. In
  \bibinfo{booktitle}{\emph{{ECOOP} 2007 - Object-Oriented Programming, 21st
  European Conference, Berlin, Germany, July 30 - August 3, 2007, Proceedings}}
  \emph{(\bibinfo{series}{Lecture Notes in Computer Science},
  Vol.~\bibinfo{volume}{4609})}, \bibfield{editor}{\bibinfo{person}{Erik
  Ernst}} (Ed.). \bibinfo{publisher}{Springer}, \bibinfo{pages}{2--27}.
\newblock
\urldef\tempurl%
\url{https://doi.org/10.1007/978-3-540-73589-2\_2}
\showDOI{\tempurl}


\bibitem[Sun et~al\mbox{.}(2018)]%
        {DBLP:conf/icse/SunLZGS18}
\bibfield{author}{\bibinfo{person}{Chengnian Sun}, \bibinfo{person}{Yuanbo Li},
  \bibinfo{person}{Qirun Zhang}, \bibinfo{person}{Tianxiao Gu}, {and}
  \bibinfo{person}{Zhendong Su}.} \bibinfo{year}{2018}\natexlab{}.
\newblock \showarticletitle{Perses: syntax-guided program reduction}. In
  \bibinfo{booktitle}{\emph{Proceedings of the 40th International Conference on
  Software Engineering, {ICSE} 2018, Gothenburg, Sweden, May 27 - June 03,
  2018}}, \bibfield{editor}{\bibinfo{person}{Michel Chaudron},
  \bibinfo{person}{Ivica Crnkovic}, \bibinfo{person}{Marsha Chechik}, {and}
  \bibinfo{person}{Mark Harman}} (Eds.). \bibinfo{publisher}{{ACM}},
  \bibinfo{pages}{361--371}.
\newblock
\urldef\tempurl%
\url{https://doi.org/10.1145/3180155.3180236}
\showDOI{\tempurl}


\bibitem[Sun et~al\mbox{.}(2020)]%
        {Sun2020TreeGen}
\bibfield{author}{\bibinfo{person}{Zeyu Sun}, \bibinfo{person}{Qihao Zhu},
  \bibinfo{person}{Yingfei Xiong}, \bibinfo{person}{Yican Sun},
  \bibinfo{person}{Lili Mou}, {and} \bibinfo{person}{Lu Zhang}.}
  \bibinfo{year}{2020}\natexlab{}.
\newblock \showarticletitle{{TreeGen}: A tree-based transformer architecture
  for code generation}. In \bibinfo{booktitle}{\emph{Proceedings of the AAAI
  Conference on Artificial Intelligence}}, Vol.~\bibinfo{volume}{34}.
  \bibinfo{pages}{8984--8991}.
\newblock


\bibitem[Tufano et~al\mbox{.}(2019)]%
        {Tufano2019}
\bibfield{author}{\bibinfo{person}{Michele Tufano}, \bibinfo{person}{Jevgenija
  Pantiuchina}, \bibinfo{person}{Cody Watson}, \bibinfo{person}{Gabriele
  Bavota}, {and} \bibinfo{person}{Denys Poshyvanyk}.}
  \bibinfo{year}{2019}\natexlab{}.
\newblock \showarticletitle{On learning meaningful code changes via neural
  machine translation}. In \bibinfo{booktitle}{\emph{Proceedings of the 41st
  International Conference on Software Engineering, {ICSE} 2019, Montreal, QC,
  Canada, May 25-31, 2019}}. \bibinfo{pages}{25--36}.
\newblock
\urldef\tempurl%
\url{https://dl.acm.org/citation.cfm?id=3339509}
\showURL{%
\tempurl}


\bibitem[Vasic et~al\mbox{.}(2019)]%
        {Vasic2019}
\bibfield{author}{\bibinfo{person}{Marko Vasic}, \bibinfo{person}{Aditya
  Kanade}, \bibinfo{person}{Petros Maniatis}, \bibinfo{person}{David Bieber},
  {and} \bibinfo{person}{Rishabh Singh}.} \bibinfo{year}{2019}\natexlab{}.
\newblock \showarticletitle{Neural Program Repair by Jointly Learning to
  Localize and Repair}. In \bibinfo{booktitle}{\emph{ICLR}}.
\newblock


\bibitem[Wan et~al\mbox{.}(2022)]%
        {Wan2022}
\bibfield{author}{\bibinfo{person}{Yao Wan}, \bibinfo{person}{Wei Zhao},
  \bibinfo{person}{Hongyu Zhang}, \bibinfo{person}{Yulei Sui},
  \bibinfo{person}{Guandong Xu}, {and} \bibinfo{person}{Hai Jin}.}
  \bibinfo{year}{2022}\natexlab{}.
\newblock \showarticletitle{What Do They Capture? A Structural Analysis of
  Pre-Trained Language Models for Source Code}. In
  \bibinfo{booktitle}{\emph{Proceedings of the 44th International Conference on
  Software Engineering}} (Pittsburgh, Pennsylvania)
  \emph{(\bibinfo{series}{ICSE '22})}. \bibinfo{publisher}{Association for
  Computing Machinery}, \bibinfo{address}{New York, NY, USA},
  \bibinfo{pages}{2377–2388}.
\newblock
\showISBNx{9781450392211}
\urldef\tempurl%
\url{https://doi.org/10.1145/3510003.3510050}
\showDOI{\tempurl}


\bibitem[Wang et~al\mbox{.}(2021)]%
        {Wang2021}
\bibfield{author}{\bibinfo{person}{Yu Wang}, \bibinfo{person}{Fengjuan Gao},
  {and} \bibinfo{person}{Linzhang Wang}.} \bibinfo{year}{2021}\natexlab{}.
\newblock \showarticletitle{Demystifying Code Summarization Models}.
\newblock \bibinfo{journal}{\emph{CoRR}} (\bibinfo{year}{2021}).
\newblock
\urldef\tempurl%
\url{https://arxiv.org/abs/2102.04625}
\showURL{%
\tempurl}


\bibitem[Wong et~al\mbox{.}(2016)]%
        {wong2016survey}
\bibfield{author}{\bibinfo{person}{W~Eric Wong}, \bibinfo{person}{Ruizhi Gao},
  \bibinfo{person}{Yihao Li}, \bibinfo{person}{Rui Abreu}, {and}
  \bibinfo{person}{Franz Wotawa}.} \bibinfo{year}{2016}\natexlab{}.
\newblock \showarticletitle{A survey on software fault localization}.
\newblock \bibinfo{journal}{\emph{IEEE Transactions on Software Engineering}}
  \bibinfo{volume}{42}, \bibinfo{number}{8} (\bibinfo{year}{2016}),
  \bibinfo{pages}{707--740}.
\newblock


\bibitem[Xia et~al\mbox{.}(2023)]%
        {xia2023automated}
\bibfield{author}{\bibinfo{person}{Chunqiu~Steven Xia},
  \bibinfo{person}{Yuxiang Wei}, {and} \bibinfo{person}{Lingming Zhang}.}
  \bibinfo{year}{2023}\natexlab{}.
\newblock \showarticletitle{Automated program repair in the era of large
  pre-trained language models}. In \bibinfo{booktitle}{\emph{Proceedings of the
  45th International Conference on Software Engineering (ICSE 2023).
  Association for Computing Machinery}}.
\newblock


\bibitem[Xia and Zhang(2022)]%
        {DBLP:conf/sigsoft/XiaZ22}
\bibfield{author}{\bibinfo{person}{Chunqiu~Steven Xia} {and}
  \bibinfo{person}{Lingming Zhang}.} \bibinfo{year}{2022}\natexlab{}.
\newblock \showarticletitle{Less training, more repairing please: revisiting
  automated program repair via zero-shot learning}. In
  \bibinfo{booktitle}{\emph{Proceedings of the 30th {ACM} Joint European
  Software Engineering Conference and Symposium on the Foundations of Software
  Engineering, {ESEC/FSE} 2022, Singapore, Singapore, November 14-18, 2022}},
  \bibfield{editor}{\bibinfo{person}{Abhik Roychoudhury},
  \bibinfo{person}{Cristian Cadar}, {and} \bibinfo{person}{Miryung Kim}}
  (Eds.). \bibinfo{publisher}{{ACM}}, \bibinfo{pages}{959--971}.
\newblock
\urldef\tempurl%
\url{https://doi.org/10.1145/3540250.3549101}
\showDOI{\tempurl}


\bibitem[Xia and Zhang(2023)]%
        {xia2023keep}
\bibfield{author}{\bibinfo{person}{Chunqiu~Steven Xia} {and}
  \bibinfo{person}{Lingming Zhang}.} \bibinfo{year}{2023}\natexlab{}.
\newblock \showarticletitle{Keep the Conversation Going: Fixing 162 out of 337
  bugs for \$0.42 each using ChatGPT}.
\newblock \bibinfo{journal}{\emph{arXiv preprint arXiv:2304.00385}}
  (\bibinfo{year}{2023}).
\newblock


\bibitem[Xu et~al\mbox{.}(2016)]%
        {Xu2016}
\bibfield{author}{\bibinfo{person}{Zhaogui Xu}, \bibinfo{person}{Xiangyu
  Zhang}, \bibinfo{person}{Lin Chen}, \bibinfo{person}{Kexin Pei}, {and}
  \bibinfo{person}{Baowen Xu}.} \bibinfo{year}{2016}\natexlab{}.
\newblock \showarticletitle{Python probabilistic type inference with natural
  language support}. In \bibinfo{booktitle}{\emph{Proceedings of the 24th {ACM}
  {SIGSOFT} International Symposium on Foundations of Software Engineering,
  {FSE} 2016, Seattle, WA, USA, November 13-18, 2016}}.
  \bibinfo{pages}{607--618}.
\newblock
\urldef\tempurl%
\url{https://doi.org/10.1145/2950290.2950343}
\showDOI{\tempurl}


\bibitem[Yasunaga and Liang(2020)]%
        {Yasunaga2020}
\bibfield{author}{\bibinfo{person}{Michihiro Yasunaga} {and}
  \bibinfo{person}{Percy Liang}.} \bibinfo{year}{2020}\natexlab{}.
\newblock \showarticletitle{Graph-based, Self-Supervised Program Repair from
  Diagnostic Feedback}. In \bibinfo{booktitle}{\emph{Proceedings of the 37th
  International Conference on Machine Learning, {ICML} 2020, 13-18 July 2020,
  Virtual Event}} \emph{(\bibinfo{series}{Proceedings of Machine Learning
  Research}, Vol.~\bibinfo{volume}{119})}. \bibinfo{publisher}{{PMLR}},
  \bibinfo{pages}{10799--10808}.
\newblock
\urldef\tempurl%
\url{http://proceedings.mlr.press/v119/yasunaga20a.html}
\showURL{%
\tempurl}


\bibitem[Ye et~al\mbox{.}(2022)]%
        {Ye2022a}
\bibfield{author}{\bibinfo{person}{He Ye}, \bibinfo{person}{Matias Martinez},
  {and} \bibinfo{person}{Martin Monperrus}.} \bibinfo{year}{2022}\natexlab{}.
\newblock \showarticletitle{Neural Program Repair with Execution-based
  Backpropagation}. In \bibinfo{booktitle}{\emph{ICSE}}.
\newblock


\bibitem[Zeller(2002)]%
        {zeller2002isolating}
\bibfield{author}{\bibinfo{person}{Andreas Zeller}.}
  \bibinfo{year}{2002}\natexlab{}.
\newblock \showarticletitle{Isolating cause-effect chains from computer
  programs}.
\newblock \bibinfo{journal}{\emph{ACM SIGSOFT Software Engineering Notes}}
  \bibinfo{volume}{27}, \bibinfo{number}{6} (\bibinfo{year}{2002}),
  \bibinfo{pages}{1--10}.
\newblock


\bibitem[Zhang et~al\mbox{.}(2021)]%
        {Zhang2021bugs}
\bibfield{author}{\bibinfo{person}{Jie~M. Zhang}, \bibinfo{person}{Feng Li},
  \bibinfo{person}{Dan Hao}, \bibinfo{person}{Meng Wang}, \bibinfo{person}{Hao
  Tang}, \bibinfo{person}{Lu Zhang}, {and} \bibinfo{person}{Mark Harman}.}
  \bibinfo{year}{2021}\natexlab{}.
\newblock \showarticletitle{A Study of Bug Resolution Characteristics in
  Popular Programming Languages}.
\newblock \bibinfo{journal}{\emph{IEEE Transactions on Software Engineering}}
  \bibinfo{volume}{47}, \bibinfo{number}{12} (\bibinfo{year}{2021}),
  \bibinfo{pages}{2684--2697}.
\newblock
\urldef\tempurl%
\url{https://doi.org/10.1109/TSE.2019.2961897}
\showDOI{\tempurl}


\bibitem[Zhu et~al\mbox{.}(2021)]%
        {Zhu2021}
\bibfield{author}{\bibinfo{person}{Qihao Zhu}, \bibinfo{person}{Zeyu Sun},
  \bibinfo{person}{Yuan{-}an Xiao}, \bibinfo{person}{Wenjie Zhang},
  \bibinfo{person}{Kang Yuan}, \bibinfo{person}{Yingfei Xiong}, {and}
  \bibinfo{person}{Lu Zhang}.} \bibinfo{year}{2021}\natexlab{}.
\newblock \showarticletitle{A syntax-guided edit decoder for neural program
  repair}. In \bibinfo{booktitle}{\emph{{ESEC/FSE} '21: 29th {ACM} Joint
  European Software Engineering Conference and Symposium on the Foundations of
  Software Engineering, Athens, Greece, August 23-28, 2021}},
  \bibfield{editor}{\bibinfo{person}{Diomidis Spinellis},
  \bibinfo{person}{Georgios Gousios}, \bibinfo{person}{Marsha Chechik}, {and}
  \bibinfo{person}{Massimiliano~Di Penta}} (Eds.). \bibinfo{publisher}{{ACM}},
  \bibinfo{pages}{341--353}.
\newblock
\urldef\tempurl%
\url{https://doi.org/10.1145/3468264.3468544}
\showDOI{\tempurl}


\end{thebibliography}

\end{document}